\documentclass[prd,aps,floats]{revtex4}

\usepackage{latexsym}
\usepackage{amssymb}
\usepackage{amsmath}

\newcommand{\etal}{{\it et al.}}

\newcommand{\apjl}{Astrophys. J. Lett.}

\newcommand{\Lie}[1]{\ensuremath{\underset{#1}{\mathcal{L}}}}

\newcommand{\connE}{\ensuremath{\tilde{\nabla}}}
\newcommand{\connM}{\ensuremath{\nabla}}

\newcommand{\conngam}{\ensuremath{\tilde{\partial}}}
\newcommand{\conneta}{\ensuremath{\bar{\partial}}}

\newcommand{\conngamt}{\ensuremath{\partial}}
\newcommand{\connetat}{\ensuremath{\hat{\partial}}}

\newcommand{\metE}{\ensuremath{\tilde{g}}}
\newcommand{\metM}{\ensuremath{g}}
\newcommand{\RWmetE}{\ensuremath{\tilde{\gamma}}}
\newcommand{\RWmetM}{\ensuremath{\gamma}}
\newcommand{\etaE}{\ensuremath{\tilde{\eta}}}
\newcommand{\etaM}{\ensuremath{\eta}}
\newcommand{\permetE}{\ensuremath{\tilde{h}}}
\newcommand{\permetM}{\ensuremath{h}}

\newcommand{\volE}{\ensuremath{\sqrt{-\metE}}}
\newcommand{\volM}{\ensuremath{\sqrt{-\metM}}}

\newcommand{\aproj}[2]{\ensuremath{s}^{#1}_{\;\;#2}}
\newcommand{\proj}[2]{\ensuremath{q}^{#1}_{\;\;#2}}
\newcommand{\RWprojE}{\ensuremath{\tilde{q}}}
\newcommand{\RWprojM}{\ensuremath{q}}
\newcommand{\etaprojM}{\ensuremath{\bar{q}}}

\newcommand{\etaproj}[2]{\ensuremath{\bar{q}}^{#1}_{\;\;#2}}

\newcommand{\XiM}{\ensuremath{\Xi}}
\newcommand{\chiM}{\ensuremath{\chi}}
\newcommand{\zetaM}{\ensuremath{\zeta}}
\newcommand{\nuM}{\ensuremath{\nu}}
\newcommand{\wM}{\ensuremath{f}}
\newcommand{\XiE}{\ensuremath{\tilde{\Xi}}}
\newcommand{\chiE}{\ensuremath{\tilde{\chi}}}
\newcommand{\zetaE}{\ensuremath{\tilde{\zeta}}}
\newcommand{\nuE}{\ensuremath{\tilde{\nu}}}
\newcommand{\wE}{\ensuremath{\tilde{f}}}

\newcommand{\PsiM}{\ensuremath{\Psi}}
\newcommand{\PhiM}{\ensuremath{\Phi}}
\newcommand{\PsiE}{\ensuremath{\tilde{\Psi}}}
\newcommand{\PhiE}{\ensuremath{\tilde{\Phi}}}

\newcommand{\RiemE}{\ensuremath{\tilde{R}}}

\newcommand{\EinE}{\ensuremath{\tilde{G}}}

\begin{document}

\title{TeVeS Cosmology : Covariant formalism for the background evolution and linear perturbation theory}

\author{Constantinos Skordis}
\affiliation{Astrophysics, University of Oxford, Keble Road,  Oxford OX1 3RH, United Kingdom and \\
Perimeter Institute, Waterloo, Ontario N2L 2Y5, Canada.} 
\email{cskordis@perimeterinstitute.ca}

\date{\today}

\renewcommand{\thefootnote}{\arabic{footnote}} \setcounter{footnote}{0}

\begin{abstract}
A relativistic theory of gravity has recently been proposed by Bekenstein, where gravity is
mediated by a tensor, a vector and a scalar field, thus called TeVeS. The theory aims at
modifying gravity in such a way as to reproduce Milgrom's MOdified Newtonian Dynamics (MOND)
in the weak field, non-relativistic limit, which provides a framework to solve the missing mass
problem in galaxies without invoking dark matter. In this paper I apply a covariant approach to
 formulate the cosmological equations for this theory, for both the background and linear
perturbations. I derive the necessary perturbed equations for scalar, vector and tensor modes
without adhering to a particular gauge. Special gauges are considered in the appendix. 
\end{abstract}

\maketitle
\tableofcontents

\section{Introduction}
Bekenstein has recently proposed a relativistic theory of gravity where gravity is mediated by a tensor, a 
vector and a scalar field, thus called TeVeS~\cite{TeVeS}, aiming at explaining the missing mass problem.

The missing mass problem is the longest standing problem of modern cosmology. It spans a wide range of scales,
from galaxies to the cosmic microwave background. The problem is easy to state : the observed mass coming from
all visible matter at the scales of interest, cannot account for the Newtonian (or General Relativistic) 
gravitational force observed acting
on the same objects. The problem has a long history~\cite{miss_mass}, and manifests as discrepancies in 
the rotation curves of galaxies, motions of clusters
of galaxies, gravitational lensing, and the absence of strong damping of linear perturbations on very large
scales, to name a few. 

One could imagine that this missing mass is composed of baryons in objects other than stars,
for example jupiter size planets or brown dwarves, collectively called MACHOS, or baryonic dark matter. 
These objects cannot 
be seen because they do not emmit light of their own. However microlensing studies did not detect 
the abundunce needed for these objects to make up for the missing mass~\cite{microlens}. Moreover the 
abundances of elements
predicted by Big Bang Nucleosynthesis (BBN) give a matter density far below the needed mass density~\cite{BBN}.

The popular approach to solving the missing mass problem,
is to posit a matter component which does not interact with electromagnetic radiation
and therefore cannot be detected by observing photons at various frequencies.
Even though it cannot be seen directly, its presence
is evident from the pull of gravity. Thus one attributes the extra gravitational force observed, to 
a "dark matter" component whose abundance is required to greatly exceed the visible matter abundance. 
Dark matter candidates have been traditionally split~\cite{BondSzalay} into "hot dark matter" and "cold dark matter", although 
an intermediate possibility, namely "warm dark matter" is sometimes being considered.

The earliest possibility considered for a dark matter candidate was a massive neutrino~\cite{GZMS,CM},
since neutrinos are   particles which are known to exist as well as being very weakly interacting. 
However, massive neutrinos cannot be the dominant form of the dark matter.
If the dark matter is composed of massive neutrinos then their mass must be at 
most $30-70eV$ for reasonable values of the 
Hubble constant, if they are not to overclose the universe~\cite{GZMS}.
On the other hand the Tremaine-Gunn inequality~\cite{tremaine_gunn_bound} gives a lower bound on the neutrino mass
if neutrinos are to be bounded gravitationally within  some radius. For example for dwarf spheroidal 
galaxies,  their mass should be greater than $\sim 300-400 eV$ which is well above the cosmologically allowed mass range.
Finally the recent Mainz and Troisk experiments
from tritium beta decay, combined with neutrino oscillation experiments give an upper limit for 
the neutrino mass  of around $2.2 - 2.5eV$~\cite{neutrino_exp}. 
Massive neutrinos are therefore ruled out as a dark matter candidate capable of solving the missing mass problem.

Cold dark matter, is composed of very massive slowly moving
and weakly interacting particles. A plethora of such particles generically arises in particle physics models beyond the standard model quite
naturally. The list of canditates is very long, ranging from light particles~\cite{light_DM},  supersymmetric
particles~\cite{SUSY_DM}, to Kalutza-Klein modes~\cite{pyrgons} and many more exotic objects. This subject (see ~\cite{particle_DM} for a nice recent review)
has been studied in great depth and has been shown to agree with observations to a very good degree.
Still, the actual nature of cold dark matter is left to speculation at the present time. 

The alternative approach is to point the finger to the laws of motion, or the gravitational field itself.
This path was initially followed by Milgrom~\cite{MOND} who proposed that for accelerations smaller than some acceleration
scale $a_0$, gravity departs from Newtonian gravity (which is still valid for large accelerations),
in such a way as to explain the flat rotation curves of galaxies. It was thus dubbed 
Modified Newtonian Dynamycs (MOND). Alternatively it was shown that one can cast a modified dynamics theory into
a modified gravity theory which provides essentially the same phenomenology~\cite{AQUAL}, although it might be
possible that the two are not entirely equivalent~\cite{milgrom}.
Other modified gravity theories beyond the MOND paradigm have also been proposed~\cite{other_modified_gravity} but I
do not consider them further in this paper.

While MOND was successful as a simple phenomenological model in describing rotation curves, it had other serious problems.
Taken at face value, it violates conservation of momentum, energy and angular momentum~\cite{MOND}.
This though stems from the fact that it is not a theory but rather an empirical law. Bekenstein and Milgrom
have found a non-relativistic self-consistent realization of MOND, 
based on an aquadratic Lagrangian, thus called AQUAL~\cite{AQUAL}.
However being non-relativistic, the theory could not make clear predictions about cosmological scales, for example the
cosmic microwave background (CMB) or the formation of linear structure (the relativistic AQUAL proposed in the
same paper suffered from acausal propagation of small perturbations).

Nevertheless, even in the absence of a consistent relativistic MOND theory, several authors have 
tried to squeeze a cosmology out of MOND.
Some authors based their calculations on general relativity to conclude that MOND
is not compatible with cosmological observations~\cite{past_GR_MOND}, 
a not so robust conclusion; the correct conclusion would 
have been that a baryonic, dark matter free universe, evolving under Einstein gravity cannot
fit the cosmological observations. In other words
they have shown the missing mass problem on cosmological scales. Nevertheless, those studies showed that
relativistic MOND theories would have to give spectra very similar to dark matter cosmologies, if they
are to be considered as serious competitors. Given that relativistic MOND theories such as TeVeS have 
generically more parameters, this would tend to favour dark matter cosmologies when tests such as
bayesian evidence are performed, unless their likelihood is strongly peaked about a very 
small region in parameter space, which is still an open question.
  
Other authors~\cite{past_cosmo_MOND}  used GR to make predictions 
for early universe cosmology (for example the CMB) 
where MOND effects were argued to be negligible and then used heuristic arguments to
make predictions about the growth of linear/non-linear structure. As \cite{SMFB} have shown, 
their CMB predictions are indeed compatible with the robust TeVeS calculations
 for some range of parameters,
but can be quite different in other cases~\cite{MBFS}. The matter power spectrum predictions  exhibit
a similar behaviour: the baryonic oscillations which
were thought to be a MOND prediction, can be absent in TeVeS, just like in
dark matter cosmology. 

Lue and Starkmann~\cite{LS} assumed that Birkhoffs 
theorem would be satisfied in a relativistic
MOND theory. If that is true one can then build a cosmology out of MOND, in the same way one can derive 
the Friedmann equation in the matter era out of Newtonian gravity. What they found was that the growth rate 
of perturbations in the matter era in this MOND-like cosmology is slower than in GR cosmology with dark matter.
Therefore if MOND is to be the limit of a relativistic theory which can successfully fit cosmological 
observations, it MUST violate Birkhoff's theorem. Indeed, TeVeS theory does violate Birkhoff's 
theorem; not only  are spherically symmetric solutions not necessarily static, 
 even the static solutions are not unique~\cite{spher_sym_TeVeS}.
The bottomline is that to make robust cosmological predictions, consistent relativistic MOND theories 
are needed.

Relativistic MOND realizations were constructed to overcome the acausal propagation
of relativistic AQUAL typically based on two metrics which are conformally related via a scalar field.
Phase coupling gravitation (PCG)~\cite{jacob1,phase_coupling1,phase_coupling2} is one such example. 
However the PCG parameters that could 
provide good MOND phenomenology were ruled out by solar system tests. Moreover, just like relativistic AQUAL,
PCG cannot provide the observed gravitational lensing from visible matter alone. Part of the problem is the conformal
relation between the two metrics~\cite{light_deflection_problem1,light_deflection_problem2}. One 
can generalized this relation to a disformal one~\cite{Disformal} by including an additive tensor in
the transformation, not related to the two metrics, for example build out of the gradient of a scalar field. However it was
soon realized that any generalized scalar-tensor gravitation theory, even with a disformal relation
between the two metrics in the theory, would produce less bending 
of light than GR and thus could not be used as a basis for
relativistic MOND~\cite{ST_light_deflection}.

Sanders stratified theory~\cite{Sanders_stratified} manages to solve 
the lensing problem. Instead of using just a scalar field to disformally
relate the two metrics in the theory, a vector field is used.
 The vector field in Sanders 
stratified theory is however non-dynamical,
which contradicts the spirit of general covariance. This was solved by
Bekenstein, who made the Sanders field dynamical by including an action for it (which is a special case of 
Jacobson-Mattingly~\cite{einstein_ether} Einstein-Ether theory action). The resulting theory was called Tensor Vector Scalar (TeVeS)~\cite{TeVeS}
gravitational theory, and was shown to provide MOND and Newtonian limits in the weak field non-relativistic limit,
and was devoid of acausal propagation of perturbations. 

Ultimately, as every physical theory, TeVeS has to face astrophysical and cosmological observations on every
scale. Some strong lensing studies have already been performed~\cite{lensing_teves}. The theory was also confronted
with observations from our own galaxy~\cite{galaxy_teves}. 
Alternatively one can test the theory with large scale structure observations or the Cosmic Microwave Background,
both of which require the propagation of linear perturbations about a Friedman-Lema\^{i}tre-Robertson-Walker(FLRW)
 background. A first study in this
direction has already been carried out~\cite{SMFB}. The subject of the present work, deals with the formulation
of linear perturbation theory in TeVeS, generalizing the equations given in~\cite{SMFB} to include  
curved spatial hypersurfaces, multiple gauges and all perturbation modes (scalar, vector and tensor modes).
 This opens up a whole new series of observational tests that can be
performed on TeVeS, involving the largest scales in the universe.

This paper is organized as follows. In the first section I give a short overview of the TeVeS theory, focusing
on the action and the field equations. At the same time I introduce a somewhat simpler notation than the original
TeVeS paper~\cite{TeVeS} which can simplify both the actions and the field equations. Relativistic fluids are then
introduced which are of particular importance to cosmology. The section concludes with the transformation of
connections between the two frames associated with the theory.

In the second section, I lay down a covariant formulation of the FLRW cosmology. The
covariant equations are derived on the basis of the symmetries of the FLRW spacetime. A short discussion
of the effective Friedmann equation follows, focussing on the time variation of the effective gravitational coupling
strength, and the definition of the relative fluid densities. A specific choice
of a coordinate system relevant to calculations for Cosmic Microwave Backround (CMB) anisotropies 
and large scale structure power spectrum, is given in the end. 

The third section takes on linear perturbations about the FLRW cosmology. The relevant tensors are perturbed 
covariantly, without adhering to a particular gauge, or perturbation mode. Thus the final perturbed equations
contain scalar, vector and tensor perturbations. The use of the covariant approach is important due to the
non-uniqueness of connections, which depend on which metric is used. Since the transformation of connections
is derived in a covariant fashion, the final perturbed equations can be derived unambiguously. This section also
gives relations of the perturbed metrics in the two frames.

In the fourth section, I take the perturbed field equations derived in the third section, and split them
into irreducible parts by seperating out the scalar, vector and tensor perturbation modes. The resulting
equations are still in a gauge non-fixed form, which makes it straightforward to check that they are indeed
gauge invariant as expected. 

Finally the paper concludes in section six, where a summary is given of the results.

The reader will also find useful the appendices. In particular
appendix-\ref{app_gauges} gives the perturbed equations for scalar modes in three 
different gauges while appendix-\ref{app_ein} gives a lot of intermediate steps in the derivation of the 
perturbed Einstein tensor. These steps were not included in the main part of the paper, to make it
more readable, but are very useful to have when following the calculations.

Through out the paper I will use a signature $+2$ metric, and the curvature conventions 
of Misner,Thorne \& Wheeler~\cite{MTW}. Greek indices are 
abstract tensor indices with no respect to any coordinate system. When writing tensor components in a particular 
coordinate system, Latin indices are used for the spatial part of the 
tensor with a "hat" on the index, while $\hat{0}$ is used for the temporal part of the tensor.
I will also use units such that 
the speed of light, Planck's constant divided by $2\pi$ and Boltzmann's constant are all equal to unity.

\section{Fundamentals of TeVeS}

\subsection{Preliminaries}
TeVeS theory is a bimetric theory where gravity is mediated by a tensor field $\metE_{\mu\nu}$ with 
associated metric-compatible connection $\connE_\mu$ and well
defined inverse $\metE^{\mu\nu}$ such that $\metE^{\mu\rho}\metE_{\rho\nu} = \delta^\mu_{\;\;\nu}$,
 a timelike (dual) vector field $A_\mu$ such that  $\metE^{\mu\nu}A_\mu A_\nu = -1$,
and a scalar field $\phi$. Matter is required to obey the
weak equivalence principle, which means that there is a metric $\metM_{\mu\nu}$ with
associated metric-compatible connection $\connM_\mu$, universal
to all matter fields, such that test particles follow its geodesics. The tensor field $\metE_{\mu\nu}$ will
be called the {\it Einstein-Hilbert frame metric} (see below) while $\metM_{\mu\nu}$ the
{\it matter frame metric}. 

The relation between the four above tensor fields (when the field equations are satisfied) is
\begin{equation}
   \metM_{\mu\nu} = e^{-2\phi}\metE_{\mu\nu} - 2\sinh(2\phi)A_\mu A_\nu
   \label{eq:metric_relation}
\end{equation}
with inverse
\begin{equation}
   \metM^{\mu\nu} = e^{2\phi}\metE^{\mu\nu} + 2\sinh(2\phi)A^\mu A^\nu
   \label{eq:inv_metric_relation}
\end{equation}
where  $A^\mu = \metE^{\mu\nu}A_\nu$.

\subsection{The action principle}
The theory is based on an action $S$, which splits as $S = S_g + S_A + S_{\phi}+S_m$, where 
$S_g$,$S_A$,$S_{\phi}$ and $S_m$ are the actions for  $\metE_{\mu\nu}$,
  vector field $A_\mu$, scalar field $\phi$ and matter respectively. 

The action for  $\metE_{\mu\nu}$, $A_\mu$ and $\phi$  is most easily written in the 
Einstein-Hilbert frame, and is such that $S_g$ is of Einstein-Hilbert form
\begin{equation}
   S_g = \frac{1}{16\pi G}\int d^4x \; \volE \; \RiemE,
\end{equation}
where $\metE$ and $\RiemE$ are  the determinant and scalar curvature of $\metE_{\mu\nu}$ respectively and
$G$ is the bare gravitational constant. Due to the complicated nature of the equations, the numerical value of $G$ will not
be the measured value of Newton's constant as measured on Earth. The precice relation between them depends on 
the spherically symmetric solution which apart from depending on the arbitrary function $V$ (see below), is also
not unique~\cite{TeVeS,spher_sym_TeVeS}. 

The action for the vector field $A_\mu$ is given by
\begin{equation}
	S_A = -\frac{1}{32\pi G}  \int d^4x \; \volE \; \left[ \; K_B \; F_{\mu\nu} F^{\mu\nu}  - 2\lambda (A_\mu A^\mu + 1)\right],
\end{equation}
where $F_{\mu\nu} = 2\connE_{[\mu} A_{\nu]}$, $F^{\mu\nu} = \metE^{\mu\alpha}\metE^{\nu\beta}F_{\alpha\beta}$,
 $\lambda$ is a Lagrange multiplier ensuring the timelike constraint on $A_\mu$ and $K_B$ is a dimensionless constant.

The action for the scalar field $\phi$ is given by
\begin{equation}
    S_{\phi} = -\frac{1}{16\pi G} \int d^4x  \volE \left[ 
    \mu \left(\metE^{\mu\nu} - A^\mu A^\nu\right)\connE_\mu\phi \connE_\nu\phi +   V(\mu) \right]
\end{equation}
 where $\mu$ is a non-dynamical dimensionless scalar field,  and $V(\mu)$ is an arbitrary 
function. The arbitrary function $V$ is  related to Bekenstein's function $F$ as
\begin{equation}
V(\mu)= \frac{1}{16 \pi \ell^2} \mu^2 F(\mu) = \frac{4\pi G^2}{\ell^2}\sigma^4_B F(G\sigma^2_B) , 
\end{equation}
where $\mu = 8\pi G \sigma_B^2$, $\sigma_B$ being Bekenstein's auxiliary scalar field.
Note that I have absorbed Bekenstein's constant $k_B$ into my definition of the function $V$.
 
The matter is coupled only to the matter frame metric $\metM_{\mu\nu}$ and thus its action is 
of the form 
\begin{equation}
 S_m[\metM,\chi^A,\partial \chi^A] = \int d^4x \; \volM \; L[ \metM ,\chi^A,\partial \chi^A] 
\end{equation}
for some generic collection of matter fields $\chi^A$. 

\subsection{The field equations}
Variation of the action w.r.t the matter frame metric gives the matter stress-energy tensor as usual by
\begin{equation}
  \delta S_m = -\frac{1}{2} \int d^4x \volM \; T_{\mu\nu} \; \delta\metM^{\mu\nu}.
\end{equation}
where $T_{\mu\nu}$ is the standard matter stress-energy tensor.

Variation of the action with respect to the three gravitational fields gives the field equations 
in the Einstein-Hilbert frame.

 The field equations for $\metE_{\mu\nu}$ are given by~\footnote{One can perform the variation using 
either $\delta g^{\mu\nu}$ or $\delta g_{\mu\nu}$. At first this 
does not give the same set of equations, the difference
being  a term proportional to $A_\mu A_\nu$. Since this is of the same form as the term involving 
the Lagrange multiplier, one can always redefine the Lagrange multiplier to eliminate the different term.
This boils down to the fact that $A_\mu$ is not unit-timelike at the level of the action which means
that (\ref{eq:metric_relation}) and (\ref{eq:inv_metric_relation}) are not inverses until after the variation.
In fact the original equations derived in ~\cite{TeVeS} (which are the same as the ones presented here)
are the simplest possible given the above reparametrization of the Lagrange multiplier.
I thank J.~D.~Bekenstein for pointing this out to me.}
\begin{equation}
	 \EinE_{\mu\nu} = Y_{\mu\nu} + 8\pi G S_{\mu\nu}
\end{equation}
where $\EinE_{\mu\nu}$ is the Einstein tensor of $\metE_{\mu\nu}$
the tensors $Y_{\mu\nu}$ and $S_{\mu\nu}$ are given by
\begin{eqnarray}
   Y_{\mu\nu} &=& \mu \left[ \connE_\mu \phi \connE_\nu\phi  - 2 A^\alpha\connE_\alpha\phi A_{(\mu}\connE_{\nu)}\phi  \right]
        + \frac{1}{2}\left(\mu V' -  V\right) \metE_{\mu\nu}   
              - \lambda A_\mu A_\nu 
  \nonumber \\ 
&&        + K_B \left[F^\alpha_{\;\;\mu}F_{\alpha\nu} - \frac{1}{4}F_{\alpha\beta}F^{\alpha\beta}\metE_{\mu\nu}\right]
  \label{eq:Y_tensor}
\end{eqnarray}
and 
\begin{equation}
   S_{\mu\nu} = T_{\mu\nu} + 2(1 - e^{-4\phi})A^\lambda T_{\lambda(\mu} A_{\nu)}
   \label{eq:einstein_stress}
\end{equation}
respectively, where $V' \equiv \frac{dV}{d\mu}$.

The field equations for the vector field $A_\mu$ are
\begin{equation}
	 K_B \connE_\mu F^\mu_{\;\;\nu} = -\lambda A_\nu -
        \mu A^\mu\connE_\mu\phi \connE_\nu\phi  + 8\pi G j_\nu \label{eq:A_eq} 
\end{equation}
where the current $j_\mu$ is given by
\begin{equation}
 j_\mu = (1 - e^{-4\phi})A^\lambda T_{\lambda\mu}
 \label{eq:current}
\end{equation}
The Lagrange multiplier is not arbitrary but can be calculated by contracting (\ref{eq:A_eq}) with $A^\mu$ and is
given by
\begin{equation}
	 \lambda =  K_B A_\nu \connE_\mu F^{\mu\nu} 
             +  \mu A^\mu A^\nu \connE_\mu\phi \connE_\nu\phi - 8\pi G\;A^\mu j_\mu 
\end{equation}
Inserting the above equation in (\ref{eq:A_eq}) one gets alternative field equations for the vector field as
\begin{equation}
 \left[\delta^\alpha_{\;\;\nu} + A^\alpha A_\nu\right] \left[K_B \connE_\mu F^\mu_{\;\;\alpha}
  + \mu A^\mu\connE_\mu\phi \connE_\alpha\phi - 8\pi G \; j_\alpha \right] = 0    
	\label{eq:A_new}
\end{equation}
which do not explicitely include the Lagrange multiplier.

The field equation for the scalar field $\phi$ is 
\begin{eqnarray}
   \connE_\mu \Gamma^\mu &=& 8\pi G J
 \label{eq:Phi_eq}
\end{eqnarray}
where
\begin{equation}
  \Gamma^\mu = \mu \left(\metE^{\mu\nu} - A^\mu A^\nu\right)\connE_\nu\phi 
  \label{eq:Gamma_eq}
\end{equation}
and  where the scalar source $J$ is given by 
\begin{equation}
  J = e^{-2\phi}\left[\metM^{\mu\nu} + 2e^{-2\phi} A^\mu A^\nu\right] T_{\mu\nu}
  \label{eq:source}
\end{equation}

Apart from the field equations above, TeVeS theory has two constraint equations which are as follows.
The first constraint is nothing but the timelike constraint on the vector field,
\begin{equation}
         \metE^{\mu\nu}A_\mu A_\mu = -1.
         \label{eq:A_constraint}
\end{equation}
which is found by varying the action with respect to the Lagrange multiplier $\lambda$.
The second constraint, fixes the non-dynamical scalar field $\mu$ in terms of the other
fields of the theory. It is given by
\begin{equation}
      \left(\metE^{\mu\nu}- A^\mu A^\nu\right)\connE_\mu\phi \connE_\nu\phi  = -V' \label{eq:mu_con} 
\end{equation}
which is found by varying the action with respect to $\mu$.
The second constraint equation must be inverted to find $\mu$ as a function of 
$\metE^{\mu\nu}$, $A^\mu$ and $\phi$. Therefore the arbitrary function $V$ and its derivatives
are nothing but functions of kinetic terms for $\phi$, contracted with $\metE^{\mu\nu}$ and $A^\mu$.

\subsection{Fluids}
The stress-energy tensor of a fluid with density $\rho$, pressure $P$, velocity $u^\mu = \metM^{\mu\nu}u_\nu$ 
and shear $\Sigma_{\mu\nu}$ is
\begin{equation}
  T_{\mu\nu} = (\rho + P ) u_\mu u_\nu + P \metM_{\mu\nu} + \Sigma_{\mu\nu}.
  \label{eq:fluid_stress}
\end{equation}
The velocity vector field is normalized with respect to the matter frame metric as $\metM^{\mu\nu} u_\mu u_\nu = u^\mu u_\mu= -1$
while the shear obeys the two conditions $\metM^{\mu\nu}\Sigma_{\mu\nu} = 0$
and $u^\mu \Sigma_{\mu\nu} = 0$.

Using (\ref{eq:einstein_stress}) one obtains
 the contribution of the fluid stress-energy tensor to the generalized Einstein equations as
\begin{eqnarray}
S_{\mu\nu} &=& (\rho+P)\left[u_\mu u_\nu + 2(1-e^{-4\phi})A^\alpha u_\alpha u_{(\mu}A_{\nu)}\right] 
    + P\left[\metM_{\mu\nu} + 4\sinh(2\phi)A_\mu A_\nu\right] \nonumber \\ 
&& + \Sigma_{\mu\nu} + 2(1 - e^{-4\phi})A^\lambda\Sigma_{\lambda(\mu}A_{\nu)} 
\label{eq:fluid_einstein_stress}
\end{eqnarray}
Similarly equation (\ref{eq:current}) gives the contribution of the fluid stress-energy tensor to the vector field equations as
\begin{equation}
j_\nu = (1 - e^{-4\phi})\left[(\rho+P)A^\mu u_\mu\;u_\nu + A^\mu\Sigma_{\mu\nu}\right] 
  + 2\sinh(2\phi)P A_\nu
\label{eq:fluid_current}
\end{equation}
Finally the contribution of the fluid stress-energy tensor to the scalar field equation  is obtained from (\ref{eq:source}) as
\begin{equation}
  J =  e^{-2\phi}\bigg[P - \rho + 2 e^{-2\phi}(\rho + P)(A^\mu u_\mu)^2 
    + 2 e^{-2\phi}A^\mu A^\nu\Sigma_{\mu\nu}\bigg]
\label{eq:fluid_source}
\end{equation}

The fluid evolution equations are obtained as usual from $\connM_\mu T^\mu_{\;\;\nu} = 0$,
where $T^\mu_{\;\;\nu} = g^{\mu\rho}T_{\rho\nu}$.
Using (\ref{eq:fluid_stress}) and expanding, gives 
\begin{equation}
  u_\nu u^\mu\connM_\mu(\rho + P) + (\rho + P)u_\nu \connM_\mu u^\mu  
 + (\rho + P) u^\mu\connM_\mu u_\nu + \connM_\nu P + \connM_\mu \Sigma^\mu_{\;\;\nu} = 0
  \label{eq:fluid_cons}
\end{equation}
where $\Sigma^\mu_{\;\;\nu} = g^{\mu\rho}\Sigma_{\rho\nu}$.
Contracting (\ref{eq:fluid_cons}) with $u^\nu$ gives the energy "conservation" equation as
\begin{equation}
  u^\mu\connM_\mu\rho  + (\rho + P) \connM_\mu u^\mu = 0,
  \label{eq:fluid_rho_cons}
\end{equation}
and subtracting (\ref{eq:fluid_rho_cons}) from (\ref{eq:fluid_cons}) yields the momentum transfer equation as
\begin{equation}
 (\rho + P) u^\mu\connM_\mu u_\nu 
 +   \left(\delta^\mu_{\;\;\nu} + u_\nu u^\mu \right)\connM_\mu P + \connM_\mu \Sigma^\mu_{\;\;\nu} = 0.
 \label{eq:fluid_theta_cons}
\end{equation}

\subsection{Transformation of connections}

I conlude this section by considering  the transformation of connections from $\connE_\mu$ to $\connM_\mu$ and vice versa.
This will come out to be very useful below, particularly in linear perturbation theory. 

Consider two metrics $g_{\mu\nu}$ and $\tilde{g}_{\mu\nu}$ on a manifold $M$, 
with connections $\nabla_\mu$ and $\tilde{\nabla}_\mu$
respectively (not necessarily the two metrics of the TeVeS theory).
The connections are required to agree on scalars, i.e. $\nabla_\mu f = \tilde{\nabla}_\mu f$ for
any $f \in C^\infty(M)$. Acting on any form $u_\mu \in T^*M$, the connections are related by
\begin{equation}
  \tilde{\nabla}_\mu u_\nu = \nabla_\mu u_\nu - D^\lambda_{\;\;\mu\nu}u_\lambda
\end{equation}
where the connection tensor $D^\lambda_{\;\;\mu\nu}$ is given by
\begin{equation}
  D^\lambda_{\;\;\mu\nu} = \frac{1}{2}g^{\lambda\rho} \left(\tilde{\nabla}_\mu g_{\rho\nu} +
   \tilde{\nabla}_\nu g_{\rho\mu} - \tilde{\nabla}_\rho g_{\mu\nu}  \right)
   \label{eq:gen_connection}
\end{equation} 

Using the metric relations (\ref{eq:metric_relation}) and (\ref{eq:inv_metric_relation}) in (\ref{eq:gen_connection}) one gets 
for TeVeS 
\begin{eqnarray}
  D^\alpha_{\;\;\mu\nu} &=& 2\delta^\alpha_{\;\;(\mu}\connE_{\nu)}\phi - \left[\metM_{\mu\nu} + 2e^{2\phi}A_\mu A_\nu \right]
         \metM^{\alpha\beta} \connE_\beta\phi 
       + 4A^\alpha A_{(\mu}\connE_{\nu)}\phi \nonumber \\
  &+& (1-e^{-4\phi}) A^\alpha \connE_{(\mu} A_{\nu)}  
   + (e^{4\phi}-1)A_{(\mu}F_{\nu)}^{\;\;\alpha} 
  + 4\sinh^2(2\phi)A_{(\mu}F_{\nu)}^{\;\;\beta}A_\beta A^\alpha
    \label{eq:teves_conn}
\end{eqnarray}
whereas with respect to $\connM_\mu$,
\begin{eqnarray}
  D^\alpha_{\;\;\mu\nu} &=& 2\delta^\alpha_{\;\;(\mu}\connM_{\nu)}\phi 
     - \left[\metE_{\mu\nu} + (e^{4\phi}+1)A_\mu A_\nu \right]\metE^{\alpha\beta} \connM_\beta\phi 
     + 2(e^{4\phi}+1)A^\alpha A_{(\mu}\connM_{\nu)}\phi \nonumber \\
     &+& (e^{4\phi} - 1) A^\alpha \connM_{(\mu} A_{\nu)} 
 + (e^{4\phi}-1)A_{(\mu}F_{\nu)}^{\;\;\alpha} 
    \label{eq:teves_conn_t}
\end{eqnarray}

The above transformations can be useful in writing the field equations in purely matter-frame form, or
Einstein-frame form, although the result could be very complicated. One should not however use these
transformations to write the actions for the three gravitational fields in matter-frame form as 
the transformations were derived under the assumption that the vector field is unit-timelike which
is not valid at the level of the action.

\section{Robertson-Walker cosmology}
\label{sec_back}
In this section I consider the  evolution of the background cosmology. 
The field equations are first found in covariant form, to facilitate an easier transition to the inclusion of
linear perturbations. A special coordinate system, common to calculations in cosmology, particularly in
cosmological perturbations is introduced at the end of the section.

\subsection{Preliminaries}
\subsubsection{Covariant description of Robertson-Walker geometry : geodesic congruences, metrics and projectors}
The background spacetime is assumed to be homogeneous and isotropic meaning that 
both of the metrics are of Robertson-Walker form. This assumption permits one
 to construct a smooth congruence of timelike geodesics, which are normal to a 
hypersurface of spatial homogeneity and isotropy. The words timelike and geodesic imply a
 metric and a compatible connection. Even though
in this theory there are two different metrics (and compatible connections), this is not a problem. The homogeneity
and isotropy is a property of the manifold and not the metric, and any of the two metrics may be used to construct
such a congruence.

Let the pair $(\RWmetE_{\mu\nu},\conngam_\mu)$ be the Robertson-Walker metric and associated metric-compatible 
connection in the 
Einstein-Hilbert frame. One can then identify the vector field $A^\mu$ as the unit vector
 field tangent to the congruence of timelike geodesics mentioned above, for the pair $(\RWmetE_{\mu\nu},\conngam_\mu)$ 
(see appendix-\ref{app_proof}).
Since by construction $A^\mu$ is hypersurface orthogonal, the Frobenious theorem , geodesic equation and 
unit-timelike condition give
\begin{equation}
  F_{\mu\nu}= 0.
\end{equation}

Now the Robertson-Walker metric is conformal to the metric of a static spacetime 
foliated by spaces of constant curvature of radius $r_c$ which will be called the conformal static metric. 
Minkoswki space (which has an infinite radius of curvature) and Einstein-static space (with positive 
radius of curvature)  are two cases of
conformal static metrics. This means that one can write $\RWmetE_{\mu\nu}$ as 
\begin{equation}
  \RWmetE_{\mu\nu} = b^2 \etaE_{\mu\nu},
  \label{eq:RW_einstein}
\end{equation}
where $b$ is the scale factor in the Einstein-Hilbert frame and where $\etaE_{\mu\nu}$ is  
the conformal static metric mentioned above with $\conneta_\mu$ its  associated metric-compatible connection.
Similarly the matter frame Robertson-Walker metric $\RWmetM_{\mu\nu}$ with associated 
metric-compatible connection  $\conngamt_\mu$, can be written as
\begin{equation}
  \RWmetM_{\mu\nu} = a^2 \etaM_{\mu\nu},
  \label{eq:RW_matter}
\end{equation}
where $a$ is the scale factor in the matter frame and $\etaM_{\mu\nu}$ is also a conformal static metric with
$\connetat_\mu$ being its metric-compatible connection.

Consider now a different unit-timelike geodesic congruence of curves with respect to $(\etaM_{\mu\nu},\connetat_\mu)$. 
Let $\bar{A}^\mu$ be the tangent vector field of this congruence assumed to be Killing (this can be accommodated 
by the symmetries of the static spacetime), which by construction obeys
\begin{equation}
  \etaM_{\mu\nu} \bar{A}^\mu \bar{A}^\nu = -1,
   \label{eq:eta_constraint}
\end{equation}
and
\begin{equation}
   \bar{A}^\nu \connetat_\nu \bar{A}_\mu = 0
   \label{eq:A_bar_geo}
\end{equation}
where $\bar{A}_\mu = \etaM_{\mu\nu} \bar{A}^\nu$.
A further property of $\bar{A}^\mu$ is that it is covariantly constant
\begin{equation}
  \connetat_\mu \bar{A}_\nu = 0
  \label{eq:A_cov_const}
\end{equation}
which follows from the fact that it is Killing, geodesic and hypersurface orthogonal.
Transforming $\etaM_{\mu\nu}$ to $\RWmetE_{\mu\nu}$ one finds  that $\bar{A}^\mu$ is related to $A^\mu$ by
\begin{equation}
  A^\mu = \frac{1}{a}e^{\bar{\phi}} \bar{A}^\mu,
\end{equation}
and
\begin{equation}
  A_\mu = a e^{-\bar{\phi}} \bar{A}_\mu. 
  \label{eq:A_bar}
\end{equation}

With the help of $A^\mu$ or equivalently $\bar{A}^\mu$ one can construct two projection tensors, given by 
 $\aproj{\mu}{\nu} = -A^\mu A_\nu$  
which projects tensors along $A^\mu$ and
 $\proj{\mu}{\nu} = \delta^\mu_{\;\;\nu} + A^\mu A_\nu$
which projects tensors on the hypersurface of homogeneity and isotropy. The two projectors have the
property that $\aproj{\mu}{\nu} A^\nu = A^\mu$, $\proj{\mu}{\alpha}\proj{\alpha}{\nu}=\proj{\mu}{\nu}$, 
and $\proj{\mu}{\nu} A^\nu = 0$. Of cource in the case of a FLRW background $\proj{\mu}{\nu} = \etaproj{\mu}{\nu}$
where 
\begin{equation}
 \etaproj{\mu}{\nu} = \delta^\mu_{\;\;\nu} + \bar{A}^\mu \bar{A}_\nu 
 \label{eq:proj_spatial}
\end{equation}
whereas when perturbations are included,
$\proj{\mu}{\nu}$  will aquire a perturbed part due to the perturbations coming from $A^\mu$ (see perturbation section)
while $\etaproj{\mu}{\nu}$ is by definition unperturbed, and the two projectors will not be equal.

\subsubsection{Relations between the scale factors and between the conformal static metrics}
A further property of the FLRW spacetime is that any gradient of a scalar function will be in the
direction $A^\mu$ or equivalently $\bar{A}^\mu$. Thus letting the background value of the scalar field be $\bar{\phi}$, its gradient
 is $\conngam_\mu \bar{\phi} = -\left(\Lie{A} \bar{\phi}\right) A_\mu = -\left(\bar{A}^\nu\conngam_\nu \bar{\phi}\right) \bar{A}_\mu$.
The same holds for any other scalar field, e.g. the scale factors in the two frames, or the background density
and pressure of the fluids etc.

The spatial metric in either frame is obtained using the spatial projector (\ref{eq:proj_spatial}) acting on
the corresponding metric. Thus in the Einstein-Hilbert frame the spatial metric is 
 $\RWprojE_{\mu\nu} = \RWmetE_{\mu\nu} + A_\mu A_\nu$ whereas in the 
matter frame it is $\RWprojM_{\mu\nu} = \RWmetM_{\mu\nu} + e^{2\bar{\phi}}A_\mu A_\nu$.
Equation (\ref{eq:metric_relation}) then implies that the two are conformally related
 as $\RWprojM_{\mu\nu} = e^{-2\bar{\phi}}\RWprojE_{\mu\nu}$, which prompts the relation 
\begin{equation}
 a = b e^{-\bar{\phi}}.
\end{equation}
between the two scale factors.

Using (\ref{eq:metric_relation}), (\ref{eq:RW_einstein}), (\ref{eq:RW_matter}) and (\ref{eq:A_bar}) 
the relation between the two conformal static metrics and $\bar{A}_\mu$ is obtained as
\begin{equation}
  \etaM_{\mu\nu} = \etaE_{\mu\nu} - (1 - e^{-4\bar{\phi}})\bar{A}_\mu\bar{A}_\nu,
  \label{eq:eta_relation}
\end{equation}
with inverse
\begin{equation}
   \etaM^{\mu\nu} = \etaE^{\mu\nu} + (e^{4\bar{\phi}} - 1)\bar{A}^\mu \bar{A}^\nu.
  \label{eq:inverse_eta_relation}
\end{equation}

\subsubsection{Connections}
 Let $\bar{C}^\alpha_{\;\;\mu\nu}$ be the connection tensor for the connection 
transformation  $\conngam_\mu \rightarrow \conneta_\mu$. 
Using (\ref{eq:gen_connection}) one gets 
\begin{equation}
  \bar{C}^\alpha_{\;\;\mu\nu} = \left[-2\delta^\alpha_{\;\;(\mu} \bar{A}_{\nu)} 
      + e^{4\bar{\phi}} \etaE_{\mu\nu} \bar{A}^\alpha\right] \Lie{\bar{A}} \ln b
    \label{eq:C_bar}
\end{equation}

Let also  $\hat{C}^\alpha_{\;\;\mu\nu}$
the connection tensor which performs the transformation $\conngamt_\mu \rightarrow \connetat_\mu$.
Using (\ref{eq:gen_connection}) one gets 
\begin{equation}
  \hat{C}^\alpha_{\;\;\mu\nu} = \left[-2\delta^\alpha_{\;\;(\mu} \bar{A}_{\nu)} 
      +   \etaM_{\mu\nu} \bar{A}^\alpha\right] \Lie{\bar{A}} \ln a. 
       \label{eq:C_hat}
\end{equation}
The reader is advised not to interprete $\conngamt_\mu$ as a partial derivative. It is a connection which annihilates
the metric $\RWmetM_{\mu\nu}$, i.e. $\conngamt_\alpha \RWmetM_{\mu\nu} = 0$ without any reference to a coordinate system.

Finally consider the connection tensor $\bar{E}^\alpha_{\;\;\mu\nu}$ for the connection transformation 
   $\conneta_\mu \rightarrow \connetat_\mu$. This is the same as
 $\conneta_\mu \rightarrow \conngam_\mu \rightarrow \conngamt_\mu \rightarrow \connetat_\mu$, and 
therefore $\bar{E}^\alpha_{\;\;\mu\nu} = -\bar{C}^\alpha_{\;\;\mu\nu} +
 D^\alpha_{\;\;\mu\nu} + \hat{C}^\alpha_{\;\;\mu\nu}$, where $D^\alpha_{\;\;\mu\nu}$ is 
 the connection tensor in (\ref{eq:teves_conn_t}) adapted to the Robertson-Walker geometry. Alternatively
  $\bar{E}^\alpha_{\;\;\mu\nu}$ is obtained directly from (\ref{eq:gen_connection}), (\ref{eq:eta_relation}),
   (\ref{eq:inverse_eta_relation}) and (\ref{eq:A_cov_const}). The final expression is
\begin{equation}
  \bar{E}^\alpha_{\;\;\mu\nu} = -2\left( \bar{A}^\beta \connetat_\beta\bar{\phi}   
       \right) \bar{A}^\alpha \bar{A}_\mu \bar{A}_\nu  
\end{equation}

\subsection{The field equations}
Here I consider the field equations adapted to the symmetries of the FLRW spacetime. I do not
explicitely consider the vector field equation as it is trivially satisfied.

\subsubsection{The fluid tensors and evolution equations}
Lets first consider the fluid related variables, $J$, $j_\mu$ and $S_{\mu\nu}$.
The fluid velocity  can be expressed in terms of $\bar{A}^\mu$ as
$u_\mu = a \bar{A}_\mu$ and $u^\mu = \frac{1}{a}\bar{A}^\mu$ which give
$A^\mu u_\mu = -e^{\bar{\phi}}$. For the same reasons as above (see appendix-\ref{app_proof})
  the fluid velocity is geodesic, i.e. $u^\mu \conngamt_\mu u_\nu = 0$.
The scalar source is then given by (\ref{eq:fluid_source}) as
\begin{equation}
 \bar{J} = e^{-2\bar{\phi}}(\bar{\rho} + 3\bar{P}) 
\end{equation}
where $\bar{\rho}$ and $\bar{P}$ are the FLRW background density and pressure of the fluid.
The fluid current is given by (\ref{eq:fluid_current}) as
\begin{equation}
  \bar{j}_\mu = - 2\sinh(2\bar{\phi}) \; e^{-\phi} \;  a \; \bar{\rho} \bar{A}_\mu 
  \label{eq:fluid_RW_current}
\end{equation}
and the fluid stress-energy tensor in the Einstein-Hilbert frame as 
\begin{equation}
  \bar{S}_{\mu\nu} = a^2\left[ - (1 - 2e^{-4\phi})\bar{\rho} \bar{A}_\mu\bar{A}_\nu + 
     \bar{P} \etaprojM_{\mu\nu}\right] \label{eq:fluid_S_RW}
\end{equation} 
where $\etaprojM_{\mu\nu} = \etaM_{\mu\lambda}\etaproj{\lambda}{\nu}$.

Changing connection from $\conngamt_\mu$  to $\connetat_\mu$,
the energy conservation equation for the fluid (\ref{eq:fluid_rho_cons}) becomes
\begin{equation}
  \bar{A}^\mu\connetat_\mu\bar{\rho} + 3 \bar{\rho}( 1  + w) \bar{A}^\mu\connetat_\mu \ln a= 0.
\end{equation}
where $w$ is the equation of state parameter such that $\bar{P} = w \bar{\rho}$.
The momentum transfer equation (\ref{eq:fluid_theta_cons})  is trivially satisfied. 

\subsubsection{The constraint equation}
Consider now the constraint (\ref{eq:mu_con}) which when adapted to the symmetries of FLRW spacetime gives
\begin{equation}
  \left(\bar{A}^\mu \connetat_\mu \bar{\phi} \right)^2 = \frac{1}{2}a^2 e^{-2\bar{\phi}} V'
 \label{eq:constraint_RW}
\end{equation}
This equation is then inverted to get $\bar{\mu} = \bar{\mu}(a,\bar{\phi},\bar{A}^\mu \connetat_\mu \bar{\phi})$.

\subsubsection{The scalar field equation}
The vector field (\ref{eq:Gamma_eq}) adapted to the FLRW symmetries  can be rewritten as
$\Gamma^\mu = - \bar{\Gamma} A^\mu$ where $\bar{\Gamma} = A_\mu \Gamma^\mu = 2\bar{\mu}\frac{e^{\bar{\phi}}}{a} (\bar{A}^\rho\connetat_\rho \bar{\phi})$.
Using the above into the scalar field equation, gives
a system of two first order equations,  which are
\begin{equation}
  \bar{A}^\mu\connetat_\mu \bar{\Gamma} = -3(\bar{A}^\rho\connetat_\rho\ln b) \bar{\Gamma}  - 8\pi G a e^{-\bar{\phi}}J
\end{equation}
and
\begin{equation}
    \bar{A}^\rho\connetat_\rho \bar{\phi}  =  \frac{1}{2\bar{\mu}}ae^{-\bar{\phi}} \bar{\Gamma}
\end{equation}
The two first order equations can then be combined into a single second-order equation which is
\begin{equation}
\connetat^2\bar{\phi} = 
  \frac{1}{U}\left[ 3 \bar{\mu}(\bar{A}^\mu \connetat_\mu \ln b)(\bar{A}^\nu \connetat_\nu\bar{\phi})
   + 4\pi G a^2 e^{-2\bar{\phi}} \bar{J}\right] 
 + \left(\bar{A}^\mu \connetat_\mu \bar{\phi}\right) \left( \bar{A}^\mu \connetat_\mu \bar{\phi}
 - \bar{A}^\mu \connetat_\mu \ln a\right) 
\end{equation}
where 
\begin{equation}
   U(\mu) = \bar{\mu} +  2\frac{V'(\mu)}{V''(\mu)}
\end{equation}
Therefore to have a solution we need $U\neq 0$.

\subsubsection{The Lagrange multiplier}
Next in line comes the lagrange multiplier which is given by
\begin{equation}
 \bar{\lambda} = \frac{1}{2}\bar{\mu} V' - 16\pi G \sinh(2\bar{\phi})\bar{\rho}
 \label{eq:RW_lagrange}
\end{equation}

\subsubsection{The generalized Einstein equations}
First lets compute the tensor $\bar{Y}_{\mu\nu}$. Using (\ref{eq:RW_lagrange}) 
 in (\ref{eq:Y_tensor}) one obtains 
\begin{equation}
  \bar{Y}_{\mu\nu} = a^2\bigg\{ \frac{1}{2}e^{2\bar{\phi}}\left(\mu V' - V\right) \etaM_{\mu\nu} 
     +\Big[\cosh(2\bar{\phi})\bar{\mu}V'
   - \sinh(2\bar{\phi}) V 
       + 8\pi G \left(1 - e^{-4\bar{\phi}}\right)\bar{\rho}\Big]\bar{A}_\mu \bar{A}_\nu \bigg\}  
\end{equation}
which when combined with (\ref{eq:fluid_S_RW}) gives the right hand side of the generalized Einstein equations as
\begin{eqnarray}
  \bar{Y}_{\mu\nu} + 8\pi G \bar{S}_{\mu\nu} &=& 
 a^2\bigg\{ \left[\frac{1}{2}e^{2\bar{\phi}}\left(\mu V' - V\right) + 8\pi G \bar{P}\right]\etaM_{\mu\nu} \nonumber \\
 &+& \Big[\cosh(2\bar{\phi})\bar{\mu}V' - \sinh(2\bar{\phi}) V 
 + 8\pi G\left( e^{-4\bar{\phi}}\bar{\rho}+\bar{P}\right)\Big]\bar{A}_\mu \bar{A}_\nu \bigg\} 
  \label{eq:RHS_G}
\end{eqnarray} 

The conformal relation of the Robertson-Walker metric to the conformal static metric makes it convenient to use
conformal transformations to calculate the Einstein tensor $\bar{G}_{\mu\nu}$ for the 
Einstein-Hilbert frame metric (\ref{eq:RW_einstein}). Now the Ricci tensor of the conformal static metric is
simply $\frac{2K}{r_c^2}\etaprojM_{\mu\nu}$ where $K$ is an integer taking the value $K=0$ for a spacially flat
hypersurface (the conformal static space is Minkowski spacetime), $K=1$ for a positively curved 
spatial hypersurface (the conformal static metric is Einstein-static spacetime) and $K=-1$ for a
 negatively curved spatial hypersurface. 

Then, after performing the conformal transformation the Einstein tensor is found to be
\begin{eqnarray}
  \bar{G}_{\mu\nu} &=& \frac{K}{r_c^2} \left[3e^{-4\bar{\phi}}\bar{A}_\mu\bar{A}_\nu - \etaprojM_{\mu\nu} \right]
   -2\left[\conneta_\mu\conneta_\nu\ln b - \left(\conneta_\mu\ln b\right) \left(\conneta_\nu\ln b \right)
 - \left(\etaE^{\alpha\beta}\conneta_\alpha \conneta_\beta\ln b\right) \etaE_{\mu\nu}\right] \nonumber \\ 
  &+& \etaE^{\alpha\beta}\left(\conneta_\alpha\ln b\right) \left(\conneta_\beta\ln b\right) \etaE_{\mu\nu} . 
 \label{eq:G_bar_b}
\end{eqnarray}
Changing connection to $\connetat_\mu$  and using (\ref{eq:eta_relation}) and (\ref{eq:inverse_eta_relation}) gives
\begin{equation}
  \bar{G}_{\mu\nu} = \frac{K}{r_c^2} \left[3e^{-4\bar{\phi}}\bar{A}_\mu\bar{A}_\nu - \etaprojM_{\mu\nu} \right]
     + 3\left(\bar{A}^\alpha \connetat_\alpha\ln b\right)^2   \bar{A}_\mu\bar{A}_\nu 
  + 2 e^{4\phi}\etaprojM_{\mu\nu}\bigg[ \connetat^2\ln b 
  -  \left(\bar{A}^\beta \connetat_\beta\ln b\right) \left( 2 \bar{A}^\alpha \connetat_\alpha\phi  
     - \frac{1}{2}\bar{A}^\alpha\connetat_\alpha\ln b\right)\bigg] . 
 \label{eq:G_hat_b}
\end{equation}

Contracting (\ref{eq:G_hat_b}) and (\ref{eq:RHS_G}) with $\bar{A}^\mu \bar{A}^\nu$ gives
the generalized Friedmann equation as 
\begin{equation}
         3 (\bar{A}^\mu\conneta_\mu\ln b )^2 = a^2 \left[
  \frac{1}{2}e^{-2\bar{\phi}}\left(\mu V' + V\right) 
 + e^{-4\bar{\phi}} \left( 8\pi G \bar{\rho} - \frac{3K}{r_c^2 a^2}\right)\right],
\end{equation}
while contracting the same equations with $\etaprojM^{\mu\nu} $ gives the
generalized  Raychandhuri equation as
\begin{equation}
  2 \connetat^2\ln b -  4 \bar{A}^\alpha \connetat_\alpha\phi\bar{A}^\beta \connetat_\beta\ln b  
  - (\bar{A}^\alpha\connetat_\alpha\ln b)^2 
= a^2 \left[\frac{1}{2}e^{-2\bar{\phi}}\left(\mu V' - V\right) + e^{-4\bar{\phi}} \left(
   8\pi G \bar{P} + \frac{K}{r_c^2 a^2}\right)\right]. 
\end{equation}
Other contractions give trivially zero.

\subsection{Choosing a coordinate system}
A most convenient coordinate system that is commonly used in cosmological perturbation theory
 is the conformal synchronous coordinate system with $t$ denoting conformal time and $x^{\hat{a}}$
 the spatial coordinates. 
There are two choices regarding the frame for which the conformal static metric used above, takes the standard form. Since
the connection to matter observables is through matter frame variables, it is more convenient to use the 
matter frame conformal static metric as taking the standard form, i.e. $\etaM_{\hat{0}\hat{0}}= -1$, 
$\etaM_{\hat{0}\hat{a}} = 0$ and $\etaM_{\hat{a}\hat{b}} = \etaprojM_{\hat{a}\hat{b}}$. 
This gives the matter frame metric as  
\begin{equation}
  ds^2 = a^2\left[ -dt^2 + \etaprojM_{\hat{a}\hat{b}} dx^{\hat{a}} dx^{\hat{b}}\right]
\end{equation}
in this coordinate system. The vanishing of the Lie derivative with respect to all the Killing vectors of the 
background spacetime gives $\phi = \bar{\phi}(t)$ only,
 as well as $\bar{A}_{\hat{a}} = 0$ and $\bar{A}^{\hat{a}} = 0$. 
The $\bar{A}_{\hat{0}}$ component is found using (\ref{eq:eta_constraint}) which gives $\bar{A}_{\hat{0}} = 1$ and
$\bar{A}^{\hat{0}} = -1$.
Finally the components of $\etaE_{\mu\nu}$ are $\etaE_{\hat{0}\hat{0}} = -e^{-4\bar{\phi}}$,
   $\etaE_{\hat{0}\hat{a}}=0$ and $\etaE_{\hat{a}\hat{b}}= \etaprojM_{\hat{a}\hat{b}}$.

Adopting the covariant equations of the previous subsection in this coordinate system gives 
the constraint equation as
\begin{equation}
   \dot{\bar{\phi}}^2 =  \frac{1}{2}a^2 e^{-2\phi} V'
\end{equation}
which must be inverted to get $\bar{\mu}(a,\bar{\phi},\dot{\bar{\phi}})$.
Similarly the scalar field equation becomes
\begin{equation}
\ddot{\bar{\phi}} =  \dot{\bar{\phi}} \left(\frac{\dot{a}}{a} - \dot{\bar{\phi}}\right) 
-  \frac{1}{U}\left[ 3 \bar{\mu}\frac{\dot{b}}{b}\dot{\bar{\phi}}
   + 4\pi G a^2 e^{-2\bar{\phi}} \bar{J}\right], 
\end{equation}
the Friedmann equation gives
\begin{equation}
         3 \frac{\dot{b}^2}{b^2} = a^2\left[\frac{1}{2}e^{-2\bar{\phi}}\left(\mu V' + V\right) +
             e^{-4\bar{\phi}}\left(8\pi G\bar{\rho} - \frac{3K}{r_c^2 a^2}\right)\right] 
\end{equation}
and the Raychandhuri equation 
\begin{equation}
  - 2 \frac{\ddot{b}}{b} + \frac{\dot{b}^2}{b^2} -  4 \frac{\dot{b}}{b}\dot{\bar{\phi}}   = 
 a^2 \left[\frac{1}{2}e^{-2\bar{\phi}}\left(\mu V' - V\right)
 +  e^{-4\bar{\phi}}\left(8\pi G\bar{P} + \frac{K}{r_c^2a^2}\right)\right] 
\end{equation}
Finally the fluid evolves according to
\begin{equation}
   \dot{\bar{\rho}} + 3\frac{\dot{a}}{a}(1 + w) \bar{\rho} = 0.
\end{equation}

\subsection{Effective Friedman equation and relative densities}
The physical Hubble parameter is as usual $H = \frac{\dot{a}}{a^2}$ and after 
transforming the scalar field time derivative as 
$\dot{\bar{\phi}} = a \frac{d\bar{\phi}}{d \ln a} H$, 
the effective Friedmann equation reads 
\begin{equation}
 3 H^2 = 8\pi G_{eff} ( \bar{\rho}_\phi + \bar{\rho}  - \frac{3K}{8\pi G \; r_c^2})
\end{equation}
where the effective gravitational coupling strength is 
\begin{equation}
  G_{eff} = G\frac{e^{-4\bar{\phi}}}{(1 + \frac{d\bar{\phi}}{d\ln a})^2}
\end{equation}
and the scalar field  density $\bar{\rho}_\phi$ is
\begin{equation}
  \bar{\rho}_\phi = \frac{1}{16\pi G} e^{2\bar{\phi}}(\bar{\mu} V' + V)
\end{equation}
Note that the effective gravitational strength is time varying. 
The relative densities $\Omega_i$ for fluid $i$ are then defined as
\begin{equation}
 \Omega_i = 8 \pi G_{eff} \frac{\bar{\rho}_i}{3 H^2} = \frac{\bar{\rho}_i}{\rho_i + \bar{\rho}_\phi}
\end{equation}
The above relation can also be used to define the relative density for the scalar field, $\Omega_\phi$.

\section{Perturbation theory}

Cosmological perturbation theory dates back to the work of Lifshitz~\cite{Lif}, 
who used a coordinate based approach and worked with the synchronous gauge. Many subsequent
studies also adopted the same approach~\cite{PT_coord_gauge}. The synchronous gauge was found to
contain spurious gauge modes~\cite{PV} causing confusion in some earlier studies as to what was the
physical growing mode. Indeed some early studies identified these residual gauge modes and 
had to carefully remove them from the solutions.
The existence of residual gauge freedom in the synchronous gauge lead Gerlach and Sengupta, Bardeen 
and others~\cite{PT_GI} to construct 
gauge invariant variables which as the name implies, were devoid of unphysical gauge modes.
Covariant~\cite{PT_cov} studies of perturbation theory were initiated by Hawking and later
 developed by Ellis and Bruni and others into a fully covariant and gauge invariant theory. 
Gauge-ready approaches, were one can always choose a gauge at will,
depending on what is more appropriate numerically were also studied by Hwang and Noh~\cite{PT_gauge_ready}. 

\subsection{Perturbations of the gravitational variables}

\subsubsection{Scalar field perturbation}
The scalar field is perturbed as
\begin{equation}
  \phi = \bar{\phi} + \varphi
\end{equation}
where $\varphi$ is the scalar field perturbation.

\subsubsection{Metric perturbations}
The Einstein-Hilbert frame metric  is perturbed as
\begin{equation}
  \metE_{\mu\nu} = b^2 \left( \etaE_{\mu\nu} + \permetE_{\mu\nu} \right)
 \label{eq:h_einstein}
\end{equation}
where $\permetE_{\mu\nu}$ is the Einstein-Hilbert frame metric perturbation. The inverse metric is given by
  $\metE^{\mu\nu} = \frac{1}{b^2}\left(\etaE^{\mu\nu} - \permetE^{\mu\nu} \right)$
where $\permetE^{\mu\nu} = \etaE^{\mu\alpha} \etaE^{\nu\beta} \permetE_{\alpha\beta}$.
One changes connection from $\connE_\mu$ to $\conneta_\mu$ via 
$\connE_\mu u_\nu = \conneta_\mu u_\nu - \left(\bar{C}_{\mu\nu}^\lambda + \tilde{f}_{\mu\nu}^\lambda\right) u_\lambda$
for some form $u_\mu$,
where the connection tensor $\tilde{f}_{\mu\nu}^\lambda$ is given by
\begin{equation}
 \tilde{f}_{\mu\nu}^\lambda = \frac{1}{2}\etaE^{\lambda\rho}\left( \conneta_\mu \etaE_{\nu\rho}
  + \conneta_\nu \etaE_{\mu\rho} - \conneta_\rho \etaE_{\mu\nu}\right)
\end{equation}

Similarly the matter frame metric is perturbed as
\begin{equation}
  \metM_{\mu\nu} = a^2 \left( \etaM_{\mu\nu} + \permetM_{\mu\nu} \right)
  \label{eq:h_matter}
\end{equation}
where $\permetM_{\mu\nu}$ is the matter frame metric perturbation. The inverse metric is given by
  $\metM^{\mu\nu} = \frac{1}{a^2}\left(\etaM^{\mu\nu} - \permetM^{\mu\nu} \right)$
where $\permetM^{\mu\nu} = \etaM^{\mu\alpha} \etaM^{\nu\beta} \permetM_{\alpha\beta}$.
Connection change from $\connM_\mu$ to $\connetat_\mu$  on  any form $u_\mu$ is as
$\connM_\mu u_\nu = \connetat_\mu u_\nu - \left(\hat{C}_{\mu\nu}^\lambda + f_{\mu\nu}^\lambda\right) u_\lambda$
where the connection tensor $f_{\mu\nu}^\lambda$ is given by
\begin{equation}
 f_{\mu\nu}^\lambda = \frac{1}{2}\etaM^{\lambda\rho}\left( \connetat_\mu \etaM_{\nu\rho}
  + \connetat_\nu \etaM_{\mu\rho} - \connetat_\rho \etaM_{\mu\nu}\right)
\end{equation}

\subsubsection{Vector field perturbations}
Let the vector field perturbation be $\alpha_\mu$, defined by
perturbing the vector field as
\begin{equation}
  A_\mu = a e^{-\bar{\phi}}\left(\bar{A}_\mu + \alpha_\mu\right)
   \label{eq:A_pert}
\end{equation}
and
\begin{equation}
  A^\mu = \frac{1}{a} e^{\bar{\phi}}\left(\bar{A}^\mu + \alpha^\mu\right)
\end{equation}
where 
\begin{equation}
\alpha^\mu = e^{-4\bar{\phi}}\left(\etaE^{\mu\nu}\alpha_\nu - \permetE^{\mu\nu}\bar{A}_\nu\right).
\end{equation}

The field strength tensor $F_{\mu\nu}$ then takes the form
\begin{equation}
 F_{\mu\nu} = 2a e^{-\bar{\phi}} \bigg[ \connetat_{[\mu} \alpha_{\nu]}
     +  \bar{A}^\alpha\left(\connetat_\alpha \bar{\phi} - \connetat_\alpha \ln a\right) 
    \bar{A}_{[\mu}\alpha_{\nu]}\bigg]
\end{equation}

Now define the "electric" and "magnetic" field parts of $F_{\mu\nu}$ as $E_\mu$ and $B_{\mu\nu}$ 
which are given by 
\begin{equation}
  E_\mu = \frac{e^{\bar{\phi}}}{a}\bar{A}^\beta F_{\alpha\beta}  
        = \frac{e^{\bar{\phi}}}{a}\etaproj{\alpha}{\mu} \bar{A}^\beta F_{\alpha\beta}
\end{equation}
and
\begin{equation}
  B_{\mu\nu} = \etaproj{\alpha}{\mu} \etaproj{\beta}{\nu} F_{\alpha\beta} = -B_{\nu\mu}.
\end{equation}
The two tensors $E_\mu$ and $B_{\mu\nu}$ 
obey $\bar{A}^\mu E_\mu = \bar{A}^\mu B_{\mu\nu} = 0$, meaning that they are 
purely spatial. 

Explicitely they are given by
\begin{equation}
 E_\mu = \connetat_\mu (\bar{A}^\beta \alpha_\beta) - \bar{A}^\beta\connetat_\beta \alpha_\mu 
 +  \bar{A}^\alpha\left(\connetat_\alpha \bar{\phi} - \connetat_\alpha \ln a\right) 
  \left(\alpha_\mu + \bar{A}_\mu \bar{A}^\beta \alpha_\beta \right)
\end{equation}
and
\begin{equation}
 B_{\mu\nu} = 2a e^{-\bar{\phi}} \etaproj{\alpha}{\mu} \etaproj{\beta}{\nu} \connetat_{[\alpha} \alpha_{\beta]}  
\end{equation}
which means that 
$F_{\mu\nu} = ae^{-\bar{\phi}}\left(\bar{A}_\mu E_\nu - E_\mu \bar{A}_\nu\right) + B_{\mu\nu}$.

\subsubsection{Perturbation of the timelike vector constraint}
Lets condider the time-like constraint on $A_\mu$ which must be preserved even
after the metric and the vector field are perturbed. This gives various relations
between the metric perturbations and the vector field perturbations. 
Perturbing the constraint gives
\begin{equation}
   \metE_{\mu\nu}A^\mu A^\nu = e^{4\bar{\phi}}
   \left(\etaE_{\mu\nu} + \permetE_{\mu\nu} \right)\left(\bar{A}^\mu + \alpha^\mu\right)
\left(\bar{A}^\nu + \alpha^\nu\right)
\end{equation}
Expanding the right hand side, transforming $\etaE_{\mu\nu}$ to $\etaM_{\mu\nu}$ with (\ref{eq:eta_relation})
and then imposing the constraint with (\ref{eq:A_constraint}) and (\ref{eq:eta_constraint}) gives 
\begin{equation}
  \permetE_{\mu\nu}\bar{A}^\mu\bar{A}^\nu = -2e^{-4\bar{\phi}}\bar{A}_\mu\alpha^\mu
\end{equation}
Two more relations are
\begin{equation}
   \bar{A}^\mu\alpha_\mu = - \bar{A}_\mu \alpha^\mu
\end{equation}
and
\begin{equation}
  \permetE^{\mu\nu}\bar{A}_\mu\bar{A}_\nu = e^{8\bar{\phi}} \permetE_{\mu\nu}\bar{A}^\mu\bar{A}^\nu 
\end{equation}

\subsubsection{Relating Einstein-Hilbert and matter frame metric perturbations}
Now lets find a relation between $\permetE_{\mu\nu}$ and $\permetM_{\mu\nu}$.
Start by perturbing the metric transformation (\ref{eq:metric_relation}) which gives
\begin{equation}
  \delta \metM_{\mu\nu} = e^{-2\bar{\phi}}\delta \metE_{\mu\nu} 
 - 2\left[e^{-2\bar{\phi}}\metE_{\mu\nu} + 2\cosh(2\bar{\phi})A_\mu A_\nu\right]\varphi 
 - 4\sinh(2\bar{\phi}) A_{(\mu} \delta A_{\nu)}
\end{equation}
Using (\ref{eq:h_einstein}), (\ref{eq:h_matter}), (\ref{eq:eta_relation}) and (\ref{eq:A_pert}) gives the
required relation which is
\begin{equation}
 \permetE_{\mu\nu} = \permetM_{\mu\nu} + 2(1 - e^{-4\bar{\phi}})\bar{A}_{(\mu}\alpha_{\nu)}
 + 2\left[\etaprojM_{\mu\nu} + \bar{A}_\mu\bar{A}_\nu\right]\varphi. 
  \label{eq:h_transform}
\end{equation}
while the inverse relation is
\begin{equation}
  \permetE^{\mu\nu} = \permetM^{\mu\nu} + 2(e^{4\bar{\phi}} -1)\bar{A}^{(\mu}\alpha^{\nu)}
 + 2\left[\etaprojM^{\mu\nu} + \bar{A}^\mu\bar{A}^\nu\right]\varphi. 
  \label{eq:inv_h_transform}
\end{equation}

\subsection{Perturbations of the fluid}

\subsubsection{Perturbations of the fluid tensors}
The fluid density and pressure are perturbed as $\rho = \bar{\rho} + \delta{\rho}$ and $P = \bar{P} + \delta P$
respectively. The fluid velocity is perturbed as
\begin{equation}
  u_\mu = a\left( \bar{A}_\mu + \theta_\mu\right)
\end{equation}
and
\begin{equation}
  u^\mu = \frac{1}{a}\left( \bar{A}^\mu + \theta^\mu\right)
\end{equation}
with
\begin{equation}
  \theta^\mu = \etaM^{\mu\nu}\theta_\mu - \permetM^{\mu\nu}\bar{A}_\nu
\end{equation}
The shear $\Sigma_{\mu\nu}$ is already a perturbation and obeys the identities
$\etaM^{\mu\nu}\Sigma_{\mu\nu} = 0$ and
$\bar{A}^\mu\Sigma_{\mu\nu} = 0$.

\subsubsection{Perturbing the timelike constraint on the fluid velocity}
This is similar to the timelike constraint of the vector field. It gives
\begin{equation} 
  \permetM^{\mu\nu}\bar{A}_\mu\bar{A}_\nu =  
  \permetM_{\mu\nu} \bar{A}^\mu\bar{A}^\nu = 2\bar{A}^\mu\theta_\mu
\end{equation}
and
\begin{equation}
  \bar{A}^\mu \theta_\mu = -\bar{A}_\mu\theta^\mu
\end{equation}

Using (\ref{eq:h_transform}), one finds a relation between the fluid velocity perturbation
and the scalar and vector field perturbations as  
\begin{equation} 
  \bar{A}^\mu\theta_\mu = \bar{A}^\mu\alpha_\mu - \varphi
\end{equation}
which also gives
\begin{equation}
  A^\mu u_\mu = -e^{\bar{\phi}}(1 + \varphi)
\end{equation}

\subsubsection{Perturbing the fluid source tensors}
Using the above relations  the scalar source perturbation $\delta J$ is found to be
\begin{equation}
\delta J = e^{-2\bar{\phi}}\left[ \delta\rho + 3\delta P -2(\bar{\rho} + 3\bar{P})\varphi \right]  
\end{equation}
Perturbing the current gives
\begin{eqnarray}
 \delta j_\mu &=& ae^{-\bar{\phi}}\bigg\{-2\sinh(2\bar{\phi})\left[ (\bar{\rho} + \bar{P})\theta_\mu
      - \bar{P}\alpha_\mu + \delta\rho \bar{A}_\mu \right] \nonumber \\
&+& \left[ -(e^{2\bar{\phi}} + 3e^{-2\bar{\phi}})\bar{\rho} + 2\sinh(2\bar{\phi})\bar{P} \right]\bar{A}_\mu \varphi\bigg\}
\end{eqnarray}
Finally, perturbing the Einstein-frame stress-energy tensor yields 
\begin{eqnarray}
  \delta S_{\mu\nu} &=& a^2 \bigg\{ \bar{P}\permetE_{\mu\nu} -2\bar{P}\etaprojM_{\mu\nu}\varphi 
  - 2(1 - e^{-4\bar{\phi}})\bar{\rho} \bar{A}_{(\mu}\alpha_{\nu)}
- 2\left[(1+3e^{-4\bar{\phi}})\bar{\rho} +  e^{-4\bar{\phi}}\bar{P}\right]\bar{A}_\mu\bar{A}_\nu\varphi \nonumber \\
 &-& (1 -2e^{-4\bar{\phi}})\delta\rho \bar{A}_\mu\bar{A}_\nu + \delta P \etaprojM_{\mu\nu} 
 + 2e^{-4\bar{\phi}}(\bar{\rho} + \bar{P}) \bar{A}_{(\mu}\theta_{\nu)}  +  \hat{\Sigma}_{\mu\nu} \bigg\}
\end{eqnarray}
where $\hat{\Sigma}_{\mu\nu} = \frac{1}{a^2}\Sigma_{\mu\nu}$.

\subsection{Perturbed field equations}

\subsubsection{The perturbed fluid equations of motion}
Let the density contrast and sound speed be given as usual by $\delta = \frac{\delta\rho}{\bar{\rho}}$ and 
$C_s^2 = \frac{\delta P}{\delta\rho}$ respectively. Then the perturbed fluid equations
become
\begin{equation}
\bar{A}^\mu\connetat_\mu\delta + 3(\bar{A}^\mu\connetat_\mu\ln a)\; (C_s^2 - w)\delta + (1+w)\left(
  \connetat_\mu\theta^\mu +  \frac{1}{2}\bar{A}^\nu\connetat_\nu \permetM^\mu_{\;\;\mu}\right) = 0 
\end{equation}
and
\begin{eqnarray}
&&  \etaproj{\nu}{\mu}\bigg[ \bar{A}^\rho\connetat_\rho\theta_\nu + (1-3w)(\bar{A}^\rho\connetat_\rho\ln a)\theta_\nu 
 -\frac{1}{2}\connetat_\nu(\bar{A}^\alpha\bar{A}^\beta\permetM_{\alpha\beta})
\nonumber \\
&&   + \connetat_\nu\left(\frac{C_s^2\delta}{1+w}\right) + \frac{\bar{A}^\rho\connetat_\rho w}{1+w}\theta_\nu
+ \frac{1}{\bar{\rho} + \bar{P}}\connetat_\rho \hat{\Sigma}^\rho_{\;\;\nu} \bigg] = 0
\end{eqnarray}

\subsubsection{Perturbed Constraint equation}
The constraint equation (\ref{eq:mu_con}) yields
\begin{equation}
      \delta \mu = 2\frac{V'}{ V''} \bar{A}^\nu \alpha_\nu 
    + \frac{4e^{2\bar{\phi}}}{a^2V''} (\bar{A}^\mu \connetat_\mu\bar{\phi}) (\bar{A}^\nu\connetat_\nu \varphi)   
      \label{eq:mu_con_per} 
\end{equation}
Unlike the unperturbed case, this need not be inverted. Rather it gives 
$\delta \mu$ directly in terms of the other variables which is then used
in the relevant places in the other perturbed equations.

\subsubsection{Perturbed scalar field equation}
Following the same approach as for the background, start from (\ref{eq:Gamma_eq}), which is then 
split into the background part $\bar{\Gamma}^\mu$ found in a previous section, and
a perturbation $\gamma^\mu$, as $\Gamma^\mu = \bar{\Gamma}^\mu + \gamma^\mu$.
Then one performs the projection $\Gamma^\mu = -\Gamma A^\mu + \proj{\mu}{\nu}\Gamma^\nu$
where $\Gamma = A_\mu \Gamma^\mu$. The scalar field  $\Gamma$ is again split into
background and perturbed part 
as $\Gamma = \bar{\Gamma} + \gamma$ 
where $\gamma = ae^{-\bar{\phi}}(\alpha_\mu\bar{\Gamma}^\mu + \bar{A}_\mu\gamma^\mu)$ is the perturbed part
which makes 
  $\gamma^\mu = - \frac{e^{\bar{\phi}}}{a}\left(\gamma + \bar{\Gamma}\bar{A}^\nu\alpha_\nu\right)\bar{A}^\mu
   + \etaproj{\mu}{\nu}\gamma^\nu$. After some calculations one gets 
the perturbation  $\gamma^\mu$ to be
\begin{eqnarray}
 \gamma^\mu &=&  \frac{\bar{\mu}}{a^2}e^{2\bar{\phi}} (\bar{A}^\lambda \connetat_\lambda\bar{\phi})
 \left[  \bar{A}^\nu\permetE^\mu_{\;\;\nu} - \bar{A}^\nu\alpha_\nu \bar{A}^\mu - \alpha^\mu\right] 
 +  \frac{\mu}{a^2}\left(e^{-2\bar{\phi}}\bar{q}^{\mu\nu} - 2e^{2\bar{\phi}}\bar{A}^\mu\bar{A}^\nu\right)
       \connetat_\nu\varphi
\nonumber \\
&& - 4\frac{e^{2\bar{\phi}}}{a^2}\frac{V'}{ V''}\left[  (\bar{A}^\rho\connetat_\rho\bar{\phi}) \bar{A}^\lambda \alpha_\lambda 
   +   \bar{A}^\beta\connetat_\beta \varphi  \right]\bar{A}^\mu 
\end{eqnarray}
while its projection on the hypersurface is
\begin{equation}
\etaproj{\mu}{\nu}\gamma^\nu = \frac{\bar{\mu}}{a^2} e^{2\bar{\phi}} (\bar{A}^\lambda \connetat_\lambda\bar{\phi})
            \left[\etaproj{\mu}{\nu}\bar{A}^\rho\permetE^\nu_{\;\;\rho} - 
               \etaproj{\mu}{\nu}\alpha^\nu \right] 
  +  \frac{\bar{\mu}}{a^2}e^{-2\bar{\phi}}\etaprojM^{\lambda\nu} \connetat_\lambda\varphi 
\end{equation}
The scalar field equation is then split into two first order equations given by
\begin{eqnarray}
 \bar{A}^\mu\connetat_\mu\gamma &=& -3(\bar{A}^\rho\connetat_\rho\ln b) \gamma
  +  \frac{\bar{\mu}}{a}e^{-3\bar{\phi}}\Delta\varphi
  + \bar{\Gamma} \bigg[\connetat_\mu\left(\etaprojM^{\mu\nu}\bar{A}^\rho\permetE_{\rho\nu}\right) - 
        \frac{1}{2} e^{-4\bar{\phi}} \etaprojM^{\mu\nu} \connetat_\mu\alpha_\nu  
\nonumber \\
    &-& \frac{1}{2} \bar{A}^\lambda\connetat_\lambda\left(\etaprojM^{\mu\nu}\permetE_{\mu\nu}\right)\bigg] 
 - 8\pi G a e^{-3\bar{\phi}}\bar{\rho}\left[ (1 + 3C_s^2)\delta  - (1 + 3w)\left(\bar{A}^\nu\alpha_\nu 
       + 2\varphi\right)\right]  
\end{eqnarray}
where $\Delta = \etaprojM^{\mu\nu}\connetat_\mu \connetat_\nu$ is the Laplace-Beltrami operator
and
\begin{equation}
     \bar{A}^\nu\connetat_\nu \varphi = 
  \frac{1}{2U}ae^{-\bar{\phi}}\gamma -  (\bar{A}^\mu \connetat_\mu\bar{\phi}) \bar{A}^\nu \alpha_\nu 
\end{equation}

\subsubsection{Perturbed vector field equation}
The divergence of $F^\mu_{\;\;\nu}$ in terms of $E_\mu$ and $B_{\mu\nu}$ is
\begin{equation}
\connE_\mu F^\mu_{\;\;\nu} = \frac{e^{\bar{\phi}}}{a}\bar{A}^\mu\connetat_\mu E_\nu 
  -\frac{e^{-3\bar{\phi}}}{a}\etaprojM^{\mu\alpha} \bar{A}_\nu \connetat_\mu E_\alpha 
  + \frac{1}{b^2}\etaprojM^{\alpha\mu}\connetat_\alpha B_{\mu\nu} 
+ \frac{e^{\bar{\phi}}}{a} (\bar{A}^\mu\connetat_\mu \ln b) E_\nu
\end{equation}

Its easier to perturb the vector field equation (\ref{eq:A_new}) which does not include the lagrange multiplier. It gives
\begin{eqnarray}
&& K_B\left[ \bar{A}^\mu\connetat_\mu E_\alpha 
  + e^{-4\bar{\phi}}\etaprojM^{\mu\nu}\connetat_\mu B_{\nu\alpha} 
+  (\bar{A}^\mu\connetat_\mu \ln b) E_\alpha \right]
+\etaproj{\nu}{\alpha}\bigg\{
   \bar{\mu}(\bar{A}^\beta\connetat_\beta\bar{\phi})
  \left[\connetat_\nu\varphi + (\bar{A}^\beta\connetat_\beta\bar{\phi}) \alpha_\nu\right] 
\nonumber \\ 
&&  + 8\pi G  a^2 (1-e^{-4\bar{\phi}})
        (\bar{\rho}+\bar{P}) \left(\theta_\nu - \alpha_\nu\right) \bigg\} = 0
\end{eqnarray}
which is a first order equation for $E_\mu$. The other equation needed is a rearrangement of the
definition of $E_\mu$ as a first order equation for $\alpha_\mu$ 
\begin{equation}
\bar{A}^\beta\connetat_\beta (\etaproj{\nu}{\mu}\alpha_\nu)  =  -E_\mu 
  + \etaproj{\nu}{\mu}\left\{ \connetat_\nu (\bar{A}^\beta \alpha_\beta)  
    +  \bar{A}^\alpha\left(\connetat_\alpha \bar{\phi} - \connetat_\alpha \ln a\right) \alpha_\nu\right\}
\end{equation}

\subsubsection{Perturbed Lagrange multiplier}
The perturbed Lagrange multiplier is given by
\begin{equation}
 \delta \lambda = K_B \frac{e^{-2\bar{\phi}}}{a^2} \etaprojM^{\mu\nu}\connetat_\mu E_\nu + \frac{1}{2}V' \delta\mu 
    + 2\bar{\mu} \frac{e^{2\bar{\phi}}}{a^2} (\bar{A}^\mu\connetat_\mu\bar{\phi}) \bar{A}^\nu \connetat_\nu\varphi 
    + \bar{\mu} V' \bar{A}^\mu \alpha_\mu - 8\pi G \delta(A^\mu j_\mu)
\end{equation}
where 
\begin{equation}
  \delta (A^\mu j_\mu) = 4\cosh(2\bar{\phi}) \bar{\rho}\varphi + 2\sinh(2\bar{\phi})\delta\rho
\end{equation}
giving
\begin{eqnarray}
 \delta \lambda &=& K_B \frac{e^{-2\bar{\phi}}}{a^2}\etaprojM^{\mu\nu}\connetat_\mu E_\nu + \frac{1}{2}V' \delta\mu 
    + 2\bar{\mu} \frac{e^{2\bar{\phi}}}{a^2} (\bar{A}^\mu\connetat_\mu\bar{\phi}) \bar{A}^\nu \connetat_\nu\varphi 
    + \bar{\mu} V' \bar{A}^\mu \alpha_\mu
\nonumber \\
    &-& 8\pi G \bar{\rho} \left[4\cosh(2\bar{\phi})\varphi + 2\sinh(2\bar{\phi})\delta\right]
\end{eqnarray}

\subsubsection{Perturbed generalized Einstein  equations}
The perturbed tensor $Y_{\mu\nu}$ yields 
\begin{eqnarray}
\delta Y_{\mu\nu} &=& \frac{1}{2} b^2 \left(\bar{\mu}\bar{V}' - \bar{V}\right)\permetE_{\mu\nu} 
 - 4\bar{\mu}(\bar{A}^\alpha\connetat_\alpha\bar{\phi})\bar{A}_{(\mu}\connetat_{\nu)}\varphi
  - K_Be^{-4\bar{\phi}}\etaprojM^{\alpha\beta} \connetat_\alpha E_\beta \bar{A}_\mu\bar{A}_\nu
\nonumber \\
 &+& \left[e^{2\bar{\phi}}\bar{\mu} - (\bar{\mu} -  2\frac{V'}{V''})e^{-2\bar{\phi}}\right] 
\left[ a^2 V'\bar{A}^\alpha\alpha_\alpha + 2 e^{2\bar{\phi}}
          (\bar{A}^\alpha\connetat_\alpha\bar{\phi})(\bar{A}^\beta\connetat_\beta\varphi) \right] \bar{A}_\mu\bar{A}_\nu
 \nonumber  \\
    &+& e^{2\bar{\phi}}\bar{\mu}\left[ a^2 V'\bar{A}^\beta\alpha_\beta 
  + 2 e^{2\bar{\phi}}(\bar{A}^\beta\connetat_\beta\bar{\phi})(\bar{A}^\alpha\connetat_\alpha\varphi)\right]
      \etaM_{\mu\nu} \nonumber \\ 
   &+& 8\pi G a^2 \bar{\rho} \left\{ (1 - e^{-4\bar{\phi}})\left[ 2\bar{A}_{(\mu}\alpha_{\nu)} 
   + \bar{A}_\mu \bar{A}_\nu \; \delta \right]
   + 2 (1 + e^{-4\bar{\phi}})\varphi \bar{A}_\mu\bar{A}_\nu \right\}
\end{eqnarray}
which when combined with $S_{\mu\nu}$, gives the right hand side of the generalized Einstein equations as
\begin{eqnarray}
\delta Y_{\mu\nu} + 8\pi G\delta S_{\mu\nu}  &=& \frac{1}{2} b^2 \left(\bar{\mu}\bar{V}' - \bar{V}\right)\permetE_{\mu\nu} 
 - 4\bar{\mu}(\bar{A}^\alpha\connetat_\alpha\bar{\phi})\bar{A}_{(\mu}\connetat_{\nu)}\varphi
  - K_Be^{-4\bar{\phi}} \etaprojM^{\alpha\beta}\connetat_\alpha E_\beta \bar{A}_\mu\bar{A}_\nu 
 \nonumber \\
 &+&\left[e^{2\bar{\phi}}\bar{\mu} - (\bar{\mu} -  2\frac{V'}{V''})e^{-2\bar{\phi}}\right]
  \left[ a^2 V'\bar{A}^\alpha\alpha_\alpha
    + 2e^{2\bar{\phi}} (\bar{A}^\alpha\connetat_\alpha\bar{\phi})(\bar{A}^\beta\connetat_\beta\varphi)
      \right]\bar{A}_\mu\bar{A}_\nu \nonumber  \\
    &+& e^{2\bar{\phi}}\bar{\mu}\left[ a^2 V'\bar{A}^\beta\alpha_\beta 
        + 2 e^{2\bar{\phi}}(\bar{A}^\beta\connetat_\beta\bar{\phi})(\bar{A}^\alpha\connetat_\alpha\varphi)\right] \etaM_{\mu\nu} 
 \nonumber \\
  &+& 8\pi G a^2 \bar{\rho}\bigg\{ w\permetE_{\mu\nu} +\left(C_s^2 \delta -2w\varphi\right)\etaprojM_{\mu\nu}
  + e^{-4\bar{\phi}} \left[\delta -2(2 + w)\varphi \right]\bar{A}_\mu\bar{A}_\nu 
 \nonumber \\
      &+&  2e^{-4\bar{\phi}}(1 + w) \bar{A}_{(\mu}\theta_{\nu)}  +  \hat{\Sigma}_{\mu\nu} \bigg\}
\end{eqnarray}

Contracting with $\bar{A}^\mu\bar{A}^\nu$ gives 
\begin{eqnarray}
\bar{A}^\mu\bar{A}^\nu\left[\delta Y_{\mu\nu} + 8\pi G\delta S_{\mu\nu}\right]  &=&  a^2 e^{-2\bar{\phi}} \left[ 2\frac{V'}{V''} V' - V \right]\bar{A}^\mu\alpha_\mu
    + 2 U (\bar{A}^\alpha\connetat_\alpha\bar{\phi})(\bar{A}^\beta\connetat_\beta\varphi)  \nonumber \\
  &-& K_Be^{-4\bar{\phi}} \etaprojM^{\alpha\beta}\connetat_\alpha E_\beta
  + 8\pi G a^2 e^{-4\bar{\phi}}\bar{\rho} \left[ \delta  - 2 \varphi - 2\bar{A}^\mu\alpha_\mu\right], 
\end{eqnarray}
contracting with $\etaproj{\mu}{\alpha}\bar{A}^\nu$ yields 
\begin{eqnarray}
\etaproj{\mu}{\alpha}\bar{A}^\nu\left[\delta Y_{\mu\nu} + 8\pi G\delta S_{\mu\nu}\right] 
 &=& \frac{1}{2} b^2 \left(\bar{\mu}\bar{V}' - \bar{V}\right) \etaproj{\mu}{\alpha}\bar{A}^\nu\permetE_{\mu\nu} 
 + 2\bar{\mu}(\bar{A}^\beta\connetat_\beta\bar{\phi}) \etaproj{\mu}{\alpha}\connetat_\mu\varphi \nonumber \\
  && + 8\pi G a^2 \bar{\rho} \bigg\{ w \etaproj{\mu}{\alpha}\bar{A}^\nu \permetE_{\mu\nu} 
 - e^{-4\bar{\phi}}(1 + w) \etaproj{\mu}{\alpha}\theta_\mu \bigg\}
\end{eqnarray}
while contracting with $\etaproj{\mu}{\alpha}\etaproj{\nu}{\beta}$ gives 
\begin{eqnarray}
\etaproj{\mu}{\alpha}\etaproj{\nu}{\beta}\left[\delta Y_{\mu\nu} + 8\pi G\delta S_{\mu\nu}\right]
  &=& a^2\left[ \frac{1}{2} e^{2\bar{\phi}} \left(\bar{\mu}\bar{V}' - \bar{V}\right) 
     + 8\pi G\bar{P}\right] \etaproj{\mu}{\alpha}\etaproj{\nu}{\beta} \permetE_{\mu\nu} 
\nonumber \\
  &+& 8\pi G a^2 \bar{\rho}\left[ (C_s^2\delta  - 2w\varphi)\etaprojM_{\alpha\beta} 
        + \etaproj{\mu}{\alpha}\etaproj{\nu}{\beta} \hat{\Sigma}_{\mu\nu} \right]
\nonumber \\
   && + \bar{\mu}e^{2\bar{\phi}}\left[a^2V' \bar{A}^\rho\alpha_\rho + 2e^{2\bar{\phi}}(\bar{A}^\rho\connetat_\rho\bar{\phi})
  (\bar{A}^\lambda\connetat_\lambda\varphi) \right]\etaprojM_{\alpha\beta}
\end{eqnarray}
The above equation can be further simplified by separating it into trace and traceless parts. The trace part is found by  
contracting with
$\etaprojM^{\mu\nu}$ and is given by
\begin{eqnarray}
 \etaprojM^{\mu\nu}\left[\delta Y_{\mu\nu} + 8\pi G\delta S_{\mu\nu}\right]
  &=& a^2\left[ \frac{1}{2} e^{2\bar{\phi}} \left(\bar{\mu}\bar{V}' - \bar{V}\right) 
   + 8\pi G\bar{P}\right] \etaprojM^{\mu\nu} \permetE_{\mu\nu} 
    + 24\pi G a^2 \bar{\rho}(C_s^2\delta - 2w\varphi) 
\nonumber \\
 &&+ 3\bar{\mu}e^{2\bar{\phi}}\left[a^2V' \bar{A}^\rho\alpha_\rho + 2e^{2\bar{\phi}}(\bar{A}^\rho\connetat_\rho\bar{\phi})
  (\bar{A}^\lambda\connetat_\lambda\varphi) \right]
\end{eqnarray}
while the traceless part is 
\begin{eqnarray}
  \left[\etaproj{\mu}{\alpha}\etaproj{\nu}{\beta} - \frac{1}{3}\etaprojM^{\mu\nu}\etaprojM_{\alpha\beta}\right]
    \left[\delta Y_{\mu\nu} + 8\pi G\delta S_{\mu\nu}\right] &=&
    a^2 \left[\etaproj{\mu}{\alpha}\etaproj{\nu}{\beta} - \frac{1}{3}\etaprojM^{\mu\nu}\etaprojM_{\alpha\beta}\right]
\bigg\{ 8\pi G  \hat{\Sigma}_{\mu\nu}
\nonumber \\ 
&+& \left[ \frac{1}{2} e^{2\bar{\phi}} \left(\bar{\mu}\bar{V}' - \bar{V}\right) 
   + 8\pi G\bar{P}\right] \permetE_{\mu\nu} \bigg\}
\end{eqnarray}

Now lets turn to the left hand side of the generalized Einstein equations.
The perturbed Ricci tensor of $\etaE_{\mu\nu} + \permetE_{\mu\nu}$  is simply given by
$2\conneta_{[\lambda} \tilde{f}^\lambda_{\nu]\mu}$, which gives the perturbed Einstein tensor as
\begin{equation}
  \delta H_{\mu\nu} = 2 \conneta_{[\lambda} \tilde{f}^\lambda_{\nu]\mu} 
  - \etaE^{\alpha\beta}\conneta_{[\lambda} \tilde{f}^\lambda_{\alpha]\beta}\etaE_{\mu\nu}
  + \frac{K}{r_c^2}\left[\etaprojM^{\alpha\beta}\permetE_{\alpha\beta}\etaE_{\mu\nu} - 3\permetE_{\mu\nu}\right] 
\end{equation}
and after expanding the connection tensors
\begin{equation}
  \delta H_{\mu\nu} = \frac{1}{2}\left[ 2\etaE^{\alpha\beta} \conneta_\beta \conneta_{(\mu} \permetE_{\nu)\alpha}
   - \conneta_\mu\conneta_\nu\permetE^\alpha_{\;\;\alpha} - \conneta^2\permetE_{\mu\nu} 
   -\left(\conneta_\alpha\conneta_\beta\permetE^{\alpha\beta} 
   - \conneta^2\permetE^\alpha_{\;\;\alpha}\right)\etaE_{\mu\nu}\right]
  + \frac{K}{r_c^2}\left[\etaprojM^{\alpha\beta}\permetE_{\alpha\beta}\etaE_{\mu\nu} - 3\permetE_{\mu\nu}\right] 
\end{equation}

The perturbed Einstein tensor in the Einstein-Hilbert frame, is then obtained via a conformal transformation as 
\begin{eqnarray}
 \delta \EinE_{\mu\nu} &=& \delta H_{\mu\nu} + 2\tilde{f}^\alpha_{\mu\nu}\conneta_\alpha\ln b 
 - \etaE_{\mu\nu} \left[ 2\permetE^{\alpha\beta}\conneta_\alpha\conneta_\beta\ln b + 2\etaE^{\alpha\beta}\tilde{f}^\lambda_{\alpha\beta}\conneta_\lambda\ln b
   + \permetE^{\alpha\beta}\bar{A}_\alpha\bar{A}_\beta (\bar{A}^\rho\conneta_\rho\ln b)^2 \right] \nonumber \\ 
 && + \permetE_{\mu\nu} \left[ 2\conneta^2\ln b - e^{4\bar{\phi}} (\bar{A}^\beta \conneta_\beta\ln b)^2 \right] 
\end{eqnarray}

Changing connection to $\connetat_\mu$ (see appendix-\ref{app_ein}) and combining terms gives 
\begin{eqnarray}
 \delta \EinE_{\mu\nu} &=& \frac{K}{r_c^2}\left[\etaprojM^{\alpha\beta}\permetE_{\alpha\beta}\etaE_{\mu\nu}
    - 3\permetE_{\mu\nu}\right] 
 -\frac{1}{2}  \connetat_\mu\connetat_\nu\permetE^\alpha_{\;\;\alpha} 
 - \frac{1}{2} \etaE^{\alpha\beta}\connetat_\alpha\connetat_\beta\permetE_{\mu\nu} 
+\frac{1}{2}\left[ \etaE^{\rho\lambda}\connetat_\rho\connetat_\lambda\permetE^\alpha_{\;\;\alpha}
   -\etaE^{\alpha\rho}\etaE^{\beta\lambda}\connetat_\alpha\connetat_\beta\permetE_{\rho\lambda} \right]\etaE_{\mu\nu}
\nonumber \\
    &+& \etaE^{\alpha\beta}\connetat_\alpha\connetat_{(\mu}\permetE_{\nu)\beta}  
+ 2e^{4\bar{\phi}}\left[
 \connetat^2\bar{\phi}
-6(\bar{A}^\rho\connetat_\rho\bar{\phi})^2
 \right] \bar{A}^\alpha\bar{A}^\beta\permetE_{\alpha\beta} \bar{A}_\mu\bar{A}_\nu
 - e^{4\bar{\phi}} (\bar{A}^\rho\connetat_\rho\bar{\phi}) \bar{A}^\lambda\connetat_\lambda
  \left(\etaproj{\alpha}{\beta}\permetE^\beta_{\;\;\alpha}\right) \etaE_{\mu\nu}
\nonumber \\
&+& (\bar{A}^\rho\connetat_\rho\bar{\phi})\bigg\{
    \left[2\etaE^{\alpha\beta}\bar{A}^\lambda\connetat_\alpha\permetE_{\beta\lambda}
- \bar{A}^\beta\connetat_\beta\permetE^\alpha_{\;\;\alpha}\right]\bar{A}_\mu\bar{A}_\nu 
\nonumber \\
&+&  e^{4\bar{\phi}}\left[
\bar{A}^\lambda\connetat_\lambda\permetE_{\mu\nu}
  - 2\bar{A}^\alpha\connetat_{(\mu}\permetE_{\nu)\alpha} 
+ 2\bar{A}^\alpha\bar{A}^\beta\bar{A}_{(\mu}\connetat_{\nu)} \permetE_{\alpha\beta} \right]\bigg\}
\nonumber \\
&-&2e^{4\bar{\phi}}\bigg\{ (\bar{A}^\rho\connetat_\rho\bar{\phi})\left[
 \bar{A}^\lambda\connetat_\lambda\left(\bar{A}_\alpha\bar{A}^\beta\permetE^\alpha_{\;\;\beta}\right)
   - \bar{A}^\lambda \etaE^{\alpha\beta}\connetat_\alpha\permetE_{\lambda\beta}\right]
 -\bar{A}_\alpha\bar{A}^\beta\permetE^\alpha_{\;\;\beta} \connetat^2\bar{\phi}\bigg\} \etaE_{\mu\nu}
\nonumber \\
&+& e^{4\bar{\phi}}(\bar{A}^\lambda\connetat_\lambda\ln b)\left[
   \bar{A}^\beta\connetat_\beta\permetE_{\mu\nu}
  -  2\bar{A}^\beta\connetat_{(\mu}\permetE_{\nu)\beta}
 -  4(\bar{A}^\rho\connetat_\rho\bar{\phi})
    \bar{A}^\alpha\bar{A}^\beta\permetE_{\alpha\beta}\bar{A}_\mu\bar{A}_\nu\right]
\nonumber \\
&-&e^{4\bar{\phi}}(\bar{A}^\rho\connetat_\rho\ln b)\left[\bar{A}^\lambda\connetat_\lambda\permetE^\alpha_{\;\;\alpha} 
  -2\bar{A}^\rho\etaE^{\alpha\beta}\connetat_\alpha\permetE_{\beta\rho}
+8(\bar{A}^\rho\connetat_\rho\bar{\phi})\bar{A}_\alpha\bar{A}^\beta\permetE^\alpha_{\;\;\beta} \right] \etaE_{\mu\nu}
\nonumber \\
&+& e^{4\bar{\phi}}\left[2\connetat^2\ln b
  - 4(\bar{A}^\lambda\connetat_\lambda\ln b)(\bar{A}^\rho\connetat_\rho\bar{\phi})
 -  (\bar{A}^\lambda\connetat_\lambda\ln b)^2\right]
\left[\permetE_{\mu\nu} + \bar{A}_\alpha\bar{A}^\beta\permetE^\alpha_{\;\;\beta} \etaE_{\mu\nu}\right]
\label{eq:delta_G}
\end{eqnarray}

Now lets perform the contractions like in all the above cases. 
Contracting  (\ref{eq:delta_G}) with $\bar{A}^\mu\bar{A}^\nu$ gives 
\begin{eqnarray}
\bar{A}^\mu\bar{A}^\nu \delta \EinE_{\mu\nu} &=&
-\frac{K}{r_c^2}\left[e^{-4\bar{\phi}}\etaprojM^{\mu\nu}\permetE_{\mu\nu}
    + 3\bar{A}^\mu\bar{A}^\nu\permetE_{\mu\nu}\right]
+ \frac{1}{2}e^{-4\bar{\phi}}
 \left[ \etaprojM^{\alpha\mu}\etaprojM^{\beta\nu}\connetat_\alpha\connetat_\beta\permetE_{\mu\nu}
- \Delta\left(\etaprojM^{\alpha\beta}\permetE_{\alpha\beta}\right) \right]
\nonumber \\
 &+& (\bar{A}^\rho\connetat_\rho\ln b)\left[
 \bar{A}^\lambda\connetat_\lambda\left(\etaprojM^{\alpha\beta}\permetE_{\alpha\beta}\right) 
  -2\bar{A}^\rho\etaprojM^{\alpha\beta}\connetat_\alpha\permetE_{\beta\rho} \right] 
\end{eqnarray}
Contracting (\ref{eq:delta_G})  with $\bar{A}^\mu\etaproj{\nu}{\alpha}$ gives 
\begin{eqnarray}
\bar{A}^\mu\etaproj{\nu}{\alpha} \delta \EinE_{\mu\nu} &=& 
-\frac{3K}{r_c^2}\bar{A}^\mu\etaproj{\nu}{\alpha}\permetE_{\mu\nu}
+\frac{1}{2} \bar{A}^\mu\etaproj{\nu}{\alpha}\etaprojM^{\rho\beta}\connetat_\rho\connetat_\mu\permetE_{\nu\beta}  
 -\frac{1}{2}  \bar{A}^\mu\etaproj{\nu}{\alpha} \connetat_\mu\connetat_\nu\left(
   \etaprojM^{\rho\beta}\permetE_{\rho\beta}\right) 
+\frac{1}{2}\proj{\nu}{\alpha}\connetat_\rho \connetat_\nu \left(\etaprojM^{\rho\beta}\bar{A}^\mu\permetE_{\mu\beta} 
     \right) 
\nonumber \\
&-& \frac{1}{2}  \bar{A}^\mu\etaproj{\nu}{\alpha} \Delta\permetE_{\mu\nu} 
- (\bar{A}^\lambda\connetat_\lambda\ln b) \etaproj{\nu}{\alpha}\connetat_\nu
  \left(\bar{A}_\mu\bar{A}^\beta\permetE^\mu_{\;\;\beta}\right)
\nonumber \\
 &+& \left[2\connetat^2\ln b
  - 4(\bar{A}^\lambda\connetat_\lambda\ln b)(\bar{A}^\rho\connetat_\rho\bar{\phi})
 -  (\bar{A}^\lambda\connetat_\lambda\ln b)^2\right] \bar{A}_\mu\etaproj{\nu}{\alpha}\permetE^\mu_{\;\;\nu} 
\end{eqnarray}
and likewise contracting (\ref{eq:delta_G}) with $\etaproj{\mu}{\alpha}\etaproj{\nu}{\beta}$ yields 
\begin{eqnarray}
\etaproj{\mu}{\alpha}\etaproj{\nu}{\beta}\delta \EinE_{\mu\nu} &=& 
\etaproj{\mu}{\alpha}\etaproj{\nu}{\beta} \frac{K}{r_c^2}\left[\etaprojM^{\rho\lambda}\permetE_{\rho\lambda}
   \etaprojM_{\mu\nu} - 3\permetE_{\mu\nu}\right]
+  \etaproj{\mu}{(\alpha}\etaproj{\nu}{\beta)} \etaprojM^{\rho\lambda}\connetat_\rho\connetat_\mu\permetE_{\nu\lambda}  
- e^{4\bar{\phi}}\etaproj{\mu}{(\alpha}\etaproj{\nu}{\beta)} \bar{A}^\rho\bar{A}^\lambda
       \connetat_\rho\connetat_\mu\permetE_{\nu\lambda}  
       \nonumber \\
        &-&
 \frac{1}{2} \etaproj{\mu}{\alpha}\etaproj{\nu}{\beta}  \connetat_\mu\connetat_\nu\permetE^\rho_{\;\;\rho} 
 - \frac{1}{2}  \etaproj{\mu}{\alpha}\etaproj{\nu}{\beta}\Delta\permetE_{\mu\nu} 
 + \frac{1}{2} e^{4\bar{\phi}}
    \etaproj{\mu}{\alpha}\etaproj{\nu}{\beta}\bar{A}^\rho\bar{A}^\lambda\connetat_\rho\connetat_\lambda\permetE_{\mu\nu} 
+\frac{1}{2}\bigg[ \Delta\permetE^\mu_{\;\;\mu}
\nonumber \\
  &-& e^{4\bar{\phi}} \bar{A}^\rho\bar{A}^\lambda\connetat_\rho\connetat_\lambda
         \left(\etaprojM^{\mu\nu}\permetE_{\mu\nu}\right)
 -\etaprojM^{\mu\rho}\etaprojM^{\nu\lambda}\connetat_\mu\connetat_\nu\permetE_{\rho\lambda}
 +2 e^{4\bar{\phi}}\bar{A}^\nu\connetat_\nu\connetat_\mu
        \left(\etaprojM^{\mu\rho}\bar{A}^\lambda \permetE_{\rho\lambda}\right)
\bigg]\etaprojM_{\alpha\beta}
\nonumber \\
&+& e^{4\bar{\phi}} (\bar{A}^\rho\connetat_\rho\bar{\phi})
 \left[2\bar{A}^\lambda \etaprojM^{\mu\nu}\connetat_\mu\permetE_{\lambda\nu} 
 - \bar{A}^\lambda\connetat_\lambda \left(\etaprojM^{\mu\nu}\permetE_{\mu\nu}\right)\right] \etaprojM_{\alpha\beta}
\nonumber \\
&+& e^{4\bar{\phi}}\left[\bar{A}^\rho\connetat_\rho\bar{\phi} +  \bar{A}^\lambda\connetat_\lambda\ln b \right]\left[
 \etaproj{\mu}{\alpha}\etaproj{\nu}{\beta}  \bar{A}^\rho\connetat_\rho\permetE_{\mu\nu}
  -  2  \etaproj{\mu}{(\alpha}\etaproj{\nu}{\beta)} \bar{A}^\rho\connetat_\mu\permetE_{\nu\rho} \right]
\nonumber \\
&-& e^{4\bar{\phi}}(\bar{A}^\rho\connetat_\rho\ln b)\left[\bar{A}^\lambda\connetat_\lambda\permetE^\mu_{\;\;\mu} 
  -2\bar{A}^\rho\etaprojM^{\mu\nu}\connetat_\mu\permetE_{\nu\rho}
  +2\bar{A}^\rho\connetat_\rho\left(\bar{A}_\mu\bar{A}^\nu\permetE^\mu_{\;\;\nu}\right) \right] 
     \etaprojM_{\alpha\beta}
\nonumber \\
&+& e^{4\bar{\phi}}\bigg[2\connetat^2\ln b
  - 4(\bar{A}^\lambda\connetat_\lambda\ln b)(\bar{A}^\rho\connetat_\rho\bar{\phi})
  \nonumber \\
 &-&  (\bar{A}^\lambda\connetat_\lambda\ln b)^2\bigg]
\left[\etaproj{\mu}{\alpha}\etaproj{\nu}{\beta}\permetE_{\mu\nu}
   + \bar{A}_\mu\bar{A}^\nu\permetE^\mu_{\;\;\nu} \etaprojM_{\alpha\beta}\right]
\end{eqnarray}

Taking the trace of the above equation, by contracting with $\etaprojM^{\alpha\beta}$ yields 
\begin{eqnarray}
 \etaprojM^{\mu\nu}\delta \EinE_{\mu\nu} &=& 
  -\frac{1}{2}\etaprojM^{\mu\nu} \etaprojM^{\rho\lambda}\connetat_\rho\connetat_\mu\permetE_{\nu\lambda}  
+2 e^{4\bar{\phi}}\bar{A}^\nu\connetat_\nu\connetat_\mu 
   \left(\etaprojM^{\mu\rho}\bar{A}^\lambda \permetE_{\rho\lambda}\right)
 + \frac{1}{2}\Delta\left(\etaprojM^{\mu\nu}\permetE_{\mu\nu}\right) 
 - \Delta\left(\bar{A}_\mu\bar{A}^\nu\permetE^\mu_{\;\;\nu}\right) 
\nonumber \\
&-& e^{4\bar{\phi}} \bar{A}^\rho\bar{A}^\lambda\connetat_\rho\connetat_\lambda\left(
      \etaprojM^{\mu\nu}\permetE_{\mu\nu}\right) 
 +2e^{4\bar{\phi}} (\bar{A}^\rho\connetat_\rho\bar{\phi})
 \left[2\bar{A}^\lambda \etaprojM^{\mu\nu}\connetat_\mu\permetE_{\lambda\nu} 
 - \bar{A}^\lambda\connetat_\lambda \left(\etaprojM^{\mu\nu}\permetE_{\mu\nu}\right)\right] 
\nonumber \\
&-&e^{4\bar{\phi}}(\bar{A}^\rho\connetat_\rho\ln b)\left[
  2\bar{A}^\lambda\connetat_\lambda\left(\etaprojM^{\mu\nu}\permetE_{\mu\nu}\right) 
  +3\bar{A}^\rho\connetat_\rho\left(\bar{A}_\mu\bar{A}^\nu\permetE^\mu_{\;\;\nu}\right) 
  -4\bar{A}^\rho\etaprojM^{\mu\nu}\connetat_\mu\permetE_{\nu\rho}\right] 
\nonumber \\
&+& e^{4\bar{\phi}}\left[2\connetat^2\ln b
  - 4(\bar{A}^\lambda\connetat_\lambda\ln b)(\bar{A}^\rho\connetat_\rho\bar{\phi})
 -  (\bar{A}^\lambda\connetat_\lambda\ln b)^2\right]
\left[\etaprojM^{\mu\nu}\permetE_{\mu\nu} + 3\bar{A}_\mu\bar{A}^\nu\permetE^\mu_{\;\;\nu} \right]
\end{eqnarray}
which gives the traceless part as
\begin{eqnarray}
 \left[\etaproj{\mu}{\alpha}\etaproj{\nu}{\beta}
  -\frac{1}{3}\etaprojM^{\mu\nu}\etaprojM_{\alpha\beta}\right]\delta \EinE_{\mu\nu} 
&=&  \left[\etaproj{\mu}{\alpha}\etaproj{\nu}{\beta}-\frac{1}{3}\etaprojM^{\mu\nu}\etaprojM_{\alpha\beta}\right]
\bigg\{ -\frac{3K}{r_c^2}\permetE_{\mu\nu} + \etaprojM^{\rho\lambda}\connetat_\rho\connetat_\mu\permetE_{\nu\lambda}
  -  e^{4\bar{\phi}} \bar{A}^\rho\bar{A}^\lambda\connetat_\rho\connetat_\mu\permetE_{\nu\lambda} 
  - \frac{1}{2} \Delta\permetE_{\mu\nu}
\nonumber \\
  &-& \frac{1}{2}  \connetat_\mu\connetat_\nu\left(\etaprojM^{\rho\lambda}
       \permetE_{\rho\lambda}\right) 
 + \frac{1}{2} \connetat_\mu\connetat_\nu\left( \bar{A}_\rho\bar{A}^\lambda\permetE^\rho_{\;\;\lambda}\right) 
   + \frac{1}{2}e^{4\bar{\phi}} \bar{A}^\rho\bar{A}^\lambda\connetat_\rho\connetat_\lambda\permetE_{\mu\nu}
\nonumber \\
 &+& e^{4\bar{\phi}}\left[\bar{A}^\rho\connetat_\rho\bar{\phi} + \bar{A}^\rho\connetat_\rho\ln b\right]
  \left[\bar{A}^\lambda\connetat_\lambda \permetE_{\mu\nu} -2 \bar{A}^\lambda\connetat_\mu\permetE_{\nu\lambda}  \right]
  \nonumber \\
&+& e^{4\bar{\phi}}\left[2\connetat^2\ln b - 4(\bar{A}^\lambda\connetat_\lambda\ln b)(\bar{A}^\rho\connetat_\rho\bar{\phi})
  -  (\bar{A}^\lambda\connetat_\lambda\ln b)^2\right]\permetE_{\mu\nu}
\bigg\}
\end{eqnarray}
The above contractions are then combined with their counterparts coming from the right hand side of the generalized
Einstein equations.

\section{Irreducible decomposition of metric perturbations}

\subsection{Harmonic mode decomposition}

\subsubsection{Einstein metric decomposition}
Lets write the Einstein metric perturbation into irreducible parts. This yields 
\begin{eqnarray}
  \permetE_{\mu\nu} &=& 2e^{-4\bar{\phi}}\XiE\bar{A}_\mu\bar{A}_\nu 
        -2 \bar{A}_{(\mu}\etaproj{\alpha}{\nu)}\connetat_{\alpha}\zetaE - 2\bar{A}_{(\mu}\permetE_{\nu)} 
\nonumber \\
         &+& \frac{1}{3}\chiE\etaprojM_{\mu\nu} + \left(\etaproj{\alpha}{\mu}\etaproj{\beta}{\nu}
         - \frac{1}{3}\etaprojM_{\mu\nu}\etaprojM^{\alpha\beta}\right)\connetat_\alpha\connetat_\beta\nuE
 + 2\connetat_\alpha \wE_{(\mu} \etaproj{\alpha}{\nu)} + \chiE_{\mu\nu} 
\label{eq:einstein_metric_decomp}
\end{eqnarray}
where $\bar{A}^\mu\permetE_\mu = \bar{A}^\mu \wE_\mu = \bar{A}^\mu\chiE_{\mu\nu} = 0$.

The variables above are classified as follows: Scalar modes ($\XiE, \zetaE$,$\chiE$ and $\nuE$),
vector modes ($\permetE_\mu$ and $\wE_\mu$)  obeying
 $\etaprojM^{\mu\nu} \connetat_\mu \permetE_\nu=0$ and $\etaprojM^{\mu\nu} \connetat_\mu \wE_\nu=0$
and tensor modes $\chiE_{\mu\nu}$ obeying $\etaprojM^{\mu\nu}\chiE_{\mu\nu}=0$ and
 $\etaprojM^{\lambda\mu} \connetat_\lambda \chiE_{\mu\nu}=0$. 

\subsubsection{Vector field decomposition}
The vector field perturbation is decomposed as
\begin{equation}
  \alpha_\mu = -\XiE \bar{A}_\mu + \etaproj{\nu}{\mu}\connetat_\nu \alpha + \beta_\mu
\end{equation}
with $\bar{A}^\mu\beta_\mu = 0$.
It contains a scalar mode $\alpha$,  given by 
$\Delta \alpha =  \etaprojM^{\mu\nu} \connetat_\mu \alpha_\nu$
and two vector modes $\beta_\mu$ obeying $\etaprojM^{\mu\nu} \connetat_\mu \beta_\nu=0$.

The "electric field" is also decomposed as
\begin{equation}
  E_\mu = \etaproj{\nu}{\mu}\connetat_\nu E + \epsilon_\mu
\end{equation}
with $E$ being a scalar mode given by $\Delta E =  \etaprojM^{\mu\nu} \connetat_\mu E_\nu$
and $\epsilon_\mu$ two vector modes obeying as usual
 $\etaprojM^{\mu\nu} \connetat_\mu \epsilon_\nu = 0$.

\subsubsection{Matter metric decomposition}
The matter frame metric is decomposed in a similar way as the Einstein frame metri as
\begin{eqnarray}
  \permetM_{\mu\nu} &=& 2\XiM\bar{A}_\mu\bar{A}_\nu 
        -2 \bar{A}_{(\mu}\etaproj{\alpha}{\nu)}\connetat_{\alpha}\zetaM - 2\bar{A}_{(\mu}\permetM_{\nu)} 
\nonumber \\
         &+& \frac{1}{3}\chiM\etaprojM_{\mu\nu} + \left(\etaproj{\alpha}{\mu}\etaproj{\beta}{\nu}
         - \frac{1}{3}\etaprojM_{\mu\nu}\etaprojM^{\alpha\beta}\right)\connetat_\alpha\connetat_\beta\nuM
 + 2\connetat_\alpha \wM_{(\mu} \etaproj{\alpha}{\nu)} + \chiM_{\mu\nu} 
\label{eq:matter_metric_decomp}
\end{eqnarray}
where $\bar{A}^\mu\permetM_\mu = \bar{A}^\mu \wM_\mu = \bar{A}^\mu\chiM_{\mu\nu} = 0$.

The variables above are classified in the same way as in the Einstein frame 
as follows: Scalar modes ($\XiM, \zetaM$,$\chiM$ and $\nuM$),
vector modes ($\permetM_\mu$ and $\wM_\mu$)  obeying
 $\etaprojM^{\mu\nu} \connetat_\mu \permetM_\nu=0$ and $\etaprojM^{\mu\nu} \connetat_\mu \wM_\nu=0$
and tensor modes $\chiM_{\mu\nu}$ obeying $\etaprojM^{\mu\nu}\chiM_{\mu\nu}=0$ and
 $\etaprojM^{\lambda\mu} \connetat_\lambda \chiM_{\mu\nu}=0$. 

\subsubsection{Relations bewteen different frame variables}
The relations between Einstein and matter frame modes can be read of from (\ref{eq:einstein_metric_decomp})
 and (\ref{eq:matter_metric_decomp}) with the 
help of (\ref{eq:h_transform}). They are as follows : 
\begin{eqnarray}
   \XiE &=& \XiM + \varphi , \\
   \zetaE &=& \zetaM - \left(1 - e^{-4\bar{\phi}}\right)\alpha ,  \\
   \chiE &=& \chiM + 6\varphi ,\\
   \nuE &=& \nuM \\ 
   \permetE_\mu &=& \permetM_\mu - (1 - e^{-4\bar{\phi}})\beta_\mu ,\\
   \wE_\mu &=& \wM_\mu ,\\
   \chiE_{\mu\nu} &=& \chiM_{\mu\nu}.
\end{eqnarray}

\subsubsection{Fluid velocity field decomposition}
The vector field perturbation is decomposed as
\begin{equation}
  \theta_\mu = -\XiM \bar{A}_\mu + \etaproj{\nu}{\mu}\connetat_\nu \theta + v_\mu
\end{equation}
with $\bar{A}^\mu v_\mu = 0$.

It contains a scalar mode : $\theta$,  given by $\Delta \theta =  \etaprojM^{\mu\nu} \connetat_\mu \theta_\nu$
and two vector modes $v_\mu$ obeying $\etaprojM^{\mu\nu} \connetat_\mu v_\nu = 0$.

\subsubsection{Fluid shear decomposition}
The shear perturbation is written as
\begin{equation}
  \hat{\Sigma}_{\mu\nu} = (\bar{\rho} + \bar{P})\etaproj{\alpha}{\mu}\etaproj{\beta}{\nu}\left[
     (\connetat_\alpha\connetat_\beta 
       - \frac{1}{3}\etaprojM_{\alpha\beta}\Delta)\Sigma + 2\sigma_{(\alpha,\beta)}
 + \sigma_{\alpha\beta} \right] 
\end{equation}
with $\bar{A}^\mu\sigma_\mu = \bar{A}^\mu\sigma_{\mu\nu} = 0$.
The variable $\Sigma$ is a scalar mode, the vector field $\sigma_\mu$ a vector mode obeying
$\etaprojM^{\mu\nu} \sigma_{\mu,\nu} = 0$, while the tensor $\sigma_{\mu\nu}$ is a tensor mode
obeying as usual $\etaprojM^{\mu\nu} \sigma_{\mu\nu} = 0 $ 
and  $\etaprojM^{\mu\nu} \sigma_{\alpha\mu,\nu} =0$.

\subsection{Gauge non-fixed equations for the scalar modes}

All scalar modes can be decomposed in terms of a complete set of eigenmodes of the Laplace-Beltrami
operator. For example a variable $A$ can be written as 
$A(x^{\hat{a}}) = \int d^3k \; Y(x^{\hat{a}},k_{\hat{b}})\; \tilde{A}(k_{\hat{b}})$,
where the eigenmodes $Y(x^{\hat{a}},k_{\hat{b}})$ obey $\left(\Delta + k^2 \right)Y =0$. In the special case
of a flat hypersurface with trivial topology, the eigenmodes are simply given 
$Y  = e^{i k_{\hat{a}} x^{\hat{a}}}$ and the integral transform above is a Fourier transform.
The wavenumber $k$ takes values depending on the geometry and topology of the spatial hypersurface.
In the case of trivial topology, $k$ takes values $k = \sqrt{k_{*}^2 - \frac{K}{r_c^2}}$, where
$k_{*}$ is continuous obeying $k_{*}\ge 0$ for a flat or negatively curved spatial hypersurface, 
and $k_{*} = \frac{N}{r_c}$ where $N$ is an integer obeying $N \ge 3$ for a positively curved spatial hypersurface. Let us also choose the same coordinate system defined in section-\ref{sec_back}.

\subsubsection{Fluid equations}
The density constrast equation for scalar modes is
\begin{equation}
\dot{\delta} = - 3 \frac{\dot{a}}{a}(C_s^2 - w)\delta + (1+w)\left(
   -k^2\theta -  \frac{1}{2} \dot{\chiM} + k^2\zetaM\right)  
\end{equation}
while the momentum divergence equation is
\begin{equation}
\dot{\theta} =  - \XiM  - \frac{\dot{a}}{a}(1 -3w)\theta +\frac{C_s^2}{1+w}\delta - \frac{\dot{w}}{1+w}\theta 
    - \frac{2}{3} \left(k^2  - \frac{3K}{r_c^2}\right)\Sigma. 
\end{equation}

\subsubsection{Scalar field equation}
The two equations equivalent to the scalar field equation are
\begin{eqnarray}
 \dot{\gamma} &=&-3\frac{\dot{b}}{b} \gamma +  \frac{\bar{\mu}}{a}e^{-3\bar{\phi}}k^2\left(\varphi 
        + \dot{\bar{\phi}}\alpha\right)
  + \frac{e^{\bar{\phi}}}{a}\bar{\mu}\dot{\bar{\phi}}\left[\dot{\chiE} -2k^2\zetaE \right] 
 \nonumber \\
 &+& 8\pi G a e^{-3\bar{\phi}}\bar{\rho}\left[( 1 + 3C_s^2) \delta
        - (1 + 3w)\left(\XiE + 2\varphi\right)\right]  
\end{eqnarray}
and
\begin{equation}
   \dot{\varphi} = -\frac{1}{2U}ae^{-\bar{\phi}}\gamma -  \dot{\bar{\phi}} \XiE 
\end{equation}

\subsubsection{Vector field equation}
The scalar mode of the perturbed vector field evolve according to the two  
 first order equations 
\begin{equation}
K_B\left(\dot{E} + \frac{\dot{b}}{b}E \right) =   - \bar{\mu}\dot{\bar{\phi}}
        (\varphi -\dot{\bar{\phi}} \alpha) 
    + 8\pi G  a^2(1 - e^{-4\bar{\phi}}) (\bar{\rho}+\bar{P}) (\theta -\alpha) 
\end{equation}
and
\begin{equation}
 \dot{\alpha}  =  E - \XiE +  \left(\dot{\bar{\phi}} - \frac{\dot{a}}{a}\right)  \alpha 
\end{equation}

\subsubsection{Generalized Einstein equations}
The scalar modes of the perturbed Generalized Einstein equations yield
the Hamiltonian constraint equation
\begin{equation}
\frac{1}{3}\left(k^2 - \frac{3K}{r_c^2}\right)\left(\chiE + k^2\nuM\right) 
   + e^{4\bar\phi}\frac{\dot{b}}{b}\left[\dot{\chiE} -2k^2\zetaE
 +6\frac{\dot{b}}{b}\XiE\right] + a e^{3\varphi} \dot{\bar{\phi}} \gamma 
 - K_B k^2E 
 = 8\pi G a^2 \bar{\rho} \left[ \delta  - 2 \varphi\right]  
\end{equation}
the momentum constraint equation
\begin{equation}
-\frac{1}{3}(\dot{\chiE} + k^2\dot{\nuM}) + \frac{K}{r_c^2}\left(2\zetaE + \dot{\nuM} \right) - 2\frac{\dot{b}}{b}\XiE =
   8\pi G a^2  e^{-4\bar{\phi}}(\bar{\rho} + \bar{P}) \theta  + 2\bar{\mu}\dot{\bar{\phi}}\varphi 
\end{equation}
and the two propagation equations
\begin{eqnarray}
&&  -\ddot{\chiE} +  2k^2\left(\dot{\zetaE} + e^{-4\bar{\phi}}\XiE\right)
 - \frac{1}{3}e^{-4\bar{\phi}}\left(k^2 - \frac{3K}{r_c^2}\right)\left(\chiE +k^2\nuM\right) 
- 2 \frac{\dot{b}}{b}\left[\dot{\chiE} + 3 \dot{\XiE} - 2k^2\zetaE \right] 
\nonumber \\
 &&
  - 2\dot{\bar{\phi}} \left[ \dot{\chiE}  - 2k^2\zetaE \right]
   + 3\frac{\mu}{U}a e^{-\bar{\phi}} \dot{\bar{\phi}} \gamma
 +6\left[-2\frac{\ddot{b}}{b} + \frac{\dot{b}^2}{b^2} - 4\dot{\bar{\phi}}\frac{\dot{b}}{b} \right]\XiE
  = 24\pi G a^2e^{-4\bar{\phi}}\bar{\rho} (C_s^2\delta - 2 w \varphi) 
\end{eqnarray}
and
\begin{equation}
 \ddot{\nuE} +  e^{-4\bar{\phi}}\left[2\XiE -\frac{1}{3}\chiE - \frac{1}{3}k^2\nuE \right] +  2\dot{\zetaE}
 + 2\left[\frac{\dot{b}}{b}+\dot{\bar{\phi}}\right] \left[ \dot{\nuE} + 2 \zetaE \right]
  =    16\pi G a^2 e^{-4\bar{\phi}}(\bar{\rho} + \bar{P})\Sigma 
\end{equation}

\subsection{Gauge non-fixed equations for the vector modes}
Let $\ell_\mu$, $m_\mu$, $n_\mu$ be an orthonormal triad of dual vector fields, normalized with respect
  to $\eta_{\mu\nu}$ which give
\begin{equation}
\etaprojM_{\mu\nu} =  \ell_\mu\ell_\nu + m_\mu m_\nu + n_\mu n_\nu
\end{equation}
and together with $\bar{A}_\mu$ they form an orthonormal tetrad for the metric $\etaM_{\mu\nu}$. 

Without loss of generality let $\ell_\mu$ be the direction of propagation of plane waves. Thus
all vector modes are orthogonal to $\ell_\mu$ for example, $\ell^\mu\beta_\mu = 0$.
Each vector mode $X$ can then be decomposed  into its two polarizations :
\begin{equation}
X_\mu = X^{+} m_\mu + X^{-} n_\mu 
\end{equation}
As it turns out there is no mixing between the two polarizations. Moreover they obey identical
equations. The "$+$" and "$-$" labels can therefore be dropped  without any confusion.

Vector modes can also be decomposed in terms of a complete set of eigenmodes of the Laplace-Beltrami
operator, just like scalar modes. The spectrum of the wavenumber $k$ is modified though, to reflect 
the spin one nature of the vector modes. In this case (again for trivial topology),
$k$ takes the values $k = \sqrt{k_{*}^2 - 2\frac{K}{r_c^2}}$, where
$k_{*}$ is continuous obeying $k_{*}\ge 0$ for a flat or negatively curved spatial hypersurface, 
and $k_{*} = \frac{N}{r_c}$ where $N$ is an integer obeying $N \ge 3$ for a positively curved spatial hypersurface.

Let us now find the equations for vector modes.
\subsubsection{Fluid equation}
The vector mode fluid equation becomes
\begin{equation}
 \dot{v} = -\left[(1 - 3w)\frac{\dot{a}}{a} + \frac{\dot{w}}{1+w}\right] v 
  	- \left(k^2 - \frac{2K}{r_c^2}\right) \sigma^{(v)}
\end{equation}

\subsubsection{Vector field equation}
The two first order equations for the vector field are
\begin{equation}
\dot{\beta} = \epsilon + \left(\dot{\bar{\phi}} - \frac{\dot{a}}{a}\right)\beta
\end{equation}
and
\begin{equation}
K_B \left[\dot{\epsilon} +  \frac{\dot{b}}{b}\epsilon
    + \left(k^2 + \frac{2K}{r_c^2}\right) e^{-4\bar{\phi}}\beta\right]
 = \dot{\bar{\phi}}^2\beta + 8\pi G a^2 (1 - e^{-4\bar{\phi}})(\bar{\rho} + \bar{P})(v - \beta)
\end{equation}

\subsubsection{Generalized Einstein equations}
The vector mode momentum constraint is 
\begin{equation}
  \left(k^2 - \frac{2K}{r_c^2}\right)\left(\dot{\wE} + \permetE \right) 
= -16\pi G a^2 e^{-4\bar{\phi}} (\bar{\rho}+\bar{P}) v 
\end{equation}
and the propagation equation is 
\begin{equation}
\ddot{\wE} + \dot{\permetE} + 2\left(\frac{\dot{b}}{b} +\dot{\bar{\phi}}\right)\left(\dot{\wE} + \permetE\right) 
   = 16\pi G a^2 e^{-4\bar{\phi}} (\bar{\rho} + \bar{P}) \sigma^{(v)}
\end{equation}

\subsection{Equations for the tensor modes}
Using the orthonormal basis defined above, one can do a similar decomposition for the tensor modes into
two polarizations. The tensor mode perturbation $\chiM_{\mu\nu}$ decomposes into a basis
 which is written as symmetrized combinations of $m_\mu$ and $n_\nu$. There
are three possibilities, namely $m_\mu m_\nu$, $n_\mu n_\nu$ and $m_\mu n_\nu + m_\nu n_\mu$. However the traceless condition
on the tensor modes, implies that the coefficient of the first two must have opposite sign, hence there are
 only two independent polarizations given by
\begin{equation}
 \chiM_{\mu\nu} =H^{+} \left(m_\mu n_\nu + m_\nu n_\mu\right) + H^{\times} \left( m_\mu m_\nu - n_\mu n_\nu\right)  
\end{equation}
Since there is no mixing of polarizations again, the labeling can be dropped.

Tensor modes are again decomposed in terms of a complete set of eigenmodes of the Laplace-Beltrami
operator. The spectrum of the wavenumber $k$ is again modified, to reflect 
the spin two nature of the tensor modes. In this case (again for trivial topology),
$k$ takes the values $k = \sqrt{k_{*}^2 - 3\frac{K}{r_c^2}}$, where
$k_{*}$ is continuous obeying $k_{*}\ge 0$ for a flat or negatively curved spatial hypersurface, 
and $k_{*} = \frac{N}{r_c}$ where $N$ is an integer obeying $N \ge 3$ for a positively curved spatial hypersurface.

 The tensor modes then obey the equation
\begin{equation}
\ddot{H} + 2\left(\frac{\dot{b}}{b} +
 \dot{\bar{\phi}}\right)\dot{H} + e^{-4\bar{\phi}} \left(k^2 + \frac{2K}{r_c^2}\right)H = 
16\pi G a^2 e^{-4\bar{\phi}}\left(\bar{\rho} + \bar{P}\right) \sigma^{(T)} 
\end{equation}

\section{Summary}
I have taken a covariant approach to formulate the linear perturbation theory about 
a spatially homogeneous and isotropic spacetime. 
The covariant approach is particularly useful in theories with two metrics where there are two different
metric compatible connections, one for  each metric.

The field equations were perturbed covariantly without adhering to a particular gauge or perturbation mode.
This allows one to check explicitely that the equations are indeed invariant under infinitesimal
gauge transformations. Mode decomposition was performed covariantly, and the equations for each 
perturbation mode were found, again without assuming a particular gauge. Special gauges for scalar modes
are given in appendix \ref{app_gauges}. While I have not considered the perturbed Boltzmann equation  
for thermalized fluids, this will remain unchanged when expressed in matter-frame variables.

This completes the linear perturbation theory for Bekenstein's TeVeS theory about a FLRW cosmological
background. The equations presented here  can be used to study the  formation of linear structure and
the Cosmic Microwave Background in this theory as was initiated in~\cite{SMFB}.

\section*{Acknowledgements}
I would like to thank Jacob Bekenstein (particularly concerning variational issues in
theories with constraints), Celine Boehm, Pedro Ferreira, David Mota and Lee Smolin
 for discussions and helpful comments. 

\appendix

\section{Any unit timelike vector field is geodesic in an FLRW universe}
\label{app_proof}
In this section I prove that that any unit timelike vector field in an FLRW universe obeys the geodesic
equation.

Let $g_{\mu\nu}$ be the Robertson-Walker metric with scale factor $a$.
 Consider now a unit-timelike vector field $t^\mu$ tangent to a 
geodesic congruence of curves. As a property of Robertson-Walker metrics, 
$t^\mu$ is always orthogonal to a hypersurface of homogeneity and isotropy (see for example \cite{wald}).
(Of course at least one such vector field exists, e.g. $t_\mu = \nabla_\mu t$ for some scalar function
$t \in C^\infty M$). Let $x^\mu$, $y^\mu$ and $z^\mu$ be three unit-spacelike vector fields, which along 
with $t^\mu$ complete an orthonormal basis on $TM$. As a property of FLRW, they can be related to 
three linearly independent Killing vectors of $M$ as  $x^\mu = \frac{1}{a} \xi^\mu_{(1)}$, $y^\mu = \frac{1}{a} \xi^\mu_{(2)}$
and $z^\mu = \frac{1}{a} \xi^\mu_{(3)}$. The above mean that 
\begin{eqnarray}
  t^\mu \nabla_\mu t_\nu = t^\mu \nabla_\nu t_\mu &=& x^\mu \nabla_\nu x_\mu = 0   \label{eq:t_x_relations} \\
  x^\mu \nabla_\mu x_\nu = \left(\Lie{t} \ln a\right) t_\nu
  \label{eq:x_geodesic}
\end{eqnarray}
 where in the last relation I have used the fact that $\Lie{x} f = 0$ for any function $f \in C^\infty M$. 

Now the vector field 
\begin{equation}
  A^\mu = (1 + c^2)^{1/2} t^\mu + c x^\mu
 \label{eq:A_norm}
\end{equation}
 is also unit-timelike by construction, for
any choice of $c \in C^\infty M$. The isotropy of $M$, implies that there is no loss of generality in
(\ref{eq:A_norm}). 

Now consider $A^\mu \nabla_\mu A_\nu$. 
Using (\ref{eq:A_norm}) one gets
\[
  A^\mu \nabla_\mu A_\nu = c\left[c t_\nu  + (1 + c^2)^{1/2} x_\nu \right]\left( \Lie{t} \ln c \right)  
  + c(1 + c^2)^{1/2} \left( t^\mu\nabla_\mu x_\nu +  x^\mu\nabla_\mu t_\nu\right)
  + c^2 \left(\Lie{t} \ln a\right) t_\nu 
\]
Now consider the term $t^\mu\nabla_\mu x_\nu + x^\mu\nabla_\mu t_\nu$ which can be expanded as
\[
 t^\mu\nabla_\mu x_\nu + x^\mu\nabla_\mu t_\nu =  \left(\Lie{t} \ln a \right) x_\nu 
   + \left( y^\alpha t^\beta \nabla_\beta x_\alpha + y^\alpha x^\beta \nabla_\beta t_\alpha\right) y_\nu 
   + \left( z^\alpha t^\beta \nabla_\beta x_\alpha + z^\alpha x^\beta \nabla_\beta t_\alpha\right) z_\nu 
\]
where (\ref{eq:t_x_relations}) and (\ref{eq:x_geodesic}) have been used.
However the coefficient of $y_\mu$ in the above relation is zero, since
\begin{eqnarray}
    y^\alpha t^\beta \nabla_\beta x_\alpha + y^\alpha x^\beta \nabla_\beta t_\alpha &=&
    y^\alpha t^\beta \nabla_\beta x_\alpha + y^\alpha \Lie{x} t_\alpha  - y^\alpha t^\beta \nabla_\alpha x_\beta \nonumber \\
  &=& 2y^\alpha t^\beta \nabla_{[\beta} x_{\alpha]} \nonumber \\
  &=& 2y^\alpha t^\beta \partial_\beta \xi_\alpha^{(1)}  \nonumber \\
  &=& 0
\end{eqnarray}
The same holds for the coefficient of $z_\mu$ for the same reason and therefore
\begin{equation}
   A^\mu \nabla_\mu A_\nu =c\left[c t_\nu  + (1 + c^2)^{1/2} x_\nu \right] \Lie{t} \ln (a c )  
\end{equation}
Therefore the choice $c = \frac{c_0}{a}$ for any constant $c_0$ means that the unit-timelike vector field
$A^\mu$ given by
\begin{equation}
  A^\mu = \left[1 + \frac{c_0^2}{a^2}\right]^{1/2} t^\mu + \frac{c_0}{a} x^\mu
 \label{eq:A_norm_final}
\end{equation}
is geodesic. However any unit-timelike vector field can be related to $t^\mu$ by (\ref{eq:A_norm_final}) which completes
the proof.

\section{Gauge choices}
\label{app_gauges}

\subsection{Gauge transformations}
Consider a vector field $\xi^\mu$ generating a local one-parameter family of local diffeomorphisms (gauge
transformations). Then
under a gauge transformation, any tensor $\mathbf{T}$ transforms as
\begin{equation}
  \mathbf{T} \rightarrow \mathbf{T} + \Lie{\xi}\mathbf{T}
\end{equation}
where $\Lie{\xi}\mathbf{T}$ is the Lie derivative of $\mathbf{T}$ along $\xi^\mu$.

Lets define a new vector field $\hat{\xi}^\mu$ by $\xi^\mu = \frac{1}{a}\hat{\xi}^\mu$
and $\xi_\mu = \metM_{\mu\nu}\xi^\nu =  a\hat{\xi}_\mu$. 
Now perform a split as
\begin{equation}
  \hat{\xi}_\mu = -\xi\bar{A}_\mu + \etaproj{\nu}{\mu}\connetat_\nu\psi + \omega_\mu  
\end{equation}
where $\bar{A}^\mu \omega_\mu = 0$.
The above vector field thus consists of two scalar modes $\xi$ and $\psi$ given by
$\xi = \bar{A}^\mu\hat{\xi}_\mu$ and $\Delta \psi = \etaprojM^{\mu\nu}\connetat_\mu\hat{\xi}_\nu$
and two vector modes $\omega_\mu$ which obey $\etaprojM^{\mu\nu}\connetat_\mu \omega_\nu = 0$.

Now one can find the gauge transformations for all the perturbed variables.
The scalar field perturbation transforms as
\begin{equation}
  \varphi' = \varphi - \frac{1}{a} (\bar{A}^\mu\connetat_\mu\bar{\phi}) \xi 
\end{equation}
while the auxiliary scalar field perturbation $\gamma$ transforms as
\begin{eqnarray}
  \gamma' &=& \gamma + 2\frac{e^{\bar{\phi}}}{a^2}U\xi\left[\connetat^2\bar{\phi} 
                     - (\bar{A}^\rho\connetat_\rho\bar{\phi})^2 +(\bar{A}^\rho\connetat_\rho\bar{\phi})
                       (\bar{A}^\nu\connetat_\nu\ln a)\right]\\ 
          &=& \gamma + \left[6\bar{\mu}\frac{e^{\bar{\phi}}}{a^2}(\bar{A}^\mu\connetat_\mu\ln b)
                            (\bar{A}^\nu\connetat_\nu\bar{\phi}) + 8\pi G e^{-3\bar{\phi}}(\bar{\rho}+3\bar{P})\right]\xi 
\end{eqnarray}

The vector field perturbation transforms as
\begin{equation}
   \alpha_\mu' = \alpha_\mu  + \frac{1}{a}\left[\connetat_\mu\xi + (\bar{A}^\nu\connetat_\nu\bar{\phi})\xi\bar{A}_\mu
       \right] 
\end{equation}
which gives
\begin{equation}
  \alpha' = \alpha  + \frac{1}{a}\xi
\end{equation}
whereas $\beta_\mu$ is gauge invariant as expected.

The vector field tensor $F_{\mu\nu}$ vanishes for the background meaning that
$E$, $\epsilon_\mu$ and $B_{\mu\nu}$ are all gauge invariant.

The matter metric perturbation transforms as
\begin{equation}
 \permetM_{\mu\nu}' = \permetM_{\mu\nu} + \frac{2}{a}\left[ \connetat_{(\mu}\hat{\xi}_{\nu)}
    +  (\bar{A}^\rho\connetat_\rho\ln a)\; \bar{A}_{(\mu}\hat{\xi}_{\nu)}
    - (\bar{A}^\rho\connetat_\rho\ln a)\; \xi \; \etaM_{\mu\nu}\right]
\end{equation}
which gives
\begin{eqnarray}
  \XiM' &=& \XiM + \frac{1}{a} \bar{A}^\mu\connetat_\mu\xi \\ 
  \zetaM' &=&  \zetaM + \frac{1}{a}\left[\bar{A}^\mu\connetat_\mu\psi - (\bar{A}^\mu\connetat_\mu\ln a)\psi + \xi
   \right]\\
  \chiM' &=& \chiM - \frac{2}{a}\left[k^2 \psi   + 3(\bar{A}^\mu\connetat_\mu\ln a) \xi \right] \\ 
  \nuM' &=&  \nuM + \frac{2}{a}\psi \\
  \permetM' &=& \permetM + \frac{1}{a}\left[\bar{A}^\mu \connetat_\mu \omega - (\bar{A}^\mu\connetat_\mu\ln a)\omega\right] \\
  \wM' &=& \wM + \frac{\omega}{a}
\end{eqnarray}
where as the Einstein-frame metric perturbations transform as
\begin{eqnarray}
  \XiE' &=& \XiE + \frac{1}{a} \left[\bar{A}^\mu\connetat_\mu\xi - (\bar{A}^\mu\connetat_\mu\bar{\phi})\xi\right] \\ 
  \zetaE' &=&  \zetaE + \frac{1}{a}\left[\bar{A}^\mu\connetat_\mu\psi - (\bar{A}^\mu\connetat_\mu\ln a)\psi 
      + e^{-4\bar{\phi}} \xi \right]\\
  \chiE' &=& \chiE - \frac{2}{a}\left[k^2 \psi   + 3(\bar{A}^\mu\connetat_\mu\ln b) \xi \right] \\ 
  \nuE' &=&  \nuE + \frac{2}{a}\psi \\
  \permetE' &=& \permetE + \frac{1}{a}\left[\bar{A}^\mu \connetat_\mu \omega - (\bar{A}^\mu\connetat_\mu\ln a)\omega\right] \\
  \wE' &=& \wE + \frac{\omega}{a}
\end{eqnarray}

The Lie derivative of the fluid velocity is $\Lie{\xi} u_\mu = \connetat_\mu \xi$
and therefore the fluid velocity transforms as
\begin{equation}
  \theta_\mu' = \theta_\mu + \frac{1}{a}\connetat_\mu\xi
\end{equation}
Now both the energy density and pressure are scalars given by
   $\rho = u_\mu u_\nu T^{\mu\nu}$ and $P = \frac{1}{3} q_{\mu\nu} T^{\mu\nu}$ and so 
\begin{eqnarray}
   \delta' &=& \delta  + \frac{3}{a}(1+w) (\bar{A}^\rho\connetat_\rho\ln a)\xi\\
   \theta' &=& \theta + \frac{1}{a}\xi \\
   \frac{1}{\bar{\rho}}\delta P' &=&  \frac{1}{\bar{\rho}}\delta P + \frac{\xi}{a}\left[
  3 w(1+w)(\bar{A}^\rho\connetat_\rho\ln a) - \bar{A}^\rho\connetat_\rho w\right]      \\
   \Sigma' &=& \Sigma \\
   v' &=& v
\end{eqnarray}

Using the gauge transformations above, a lengthly calculation shows that the 
gauge non-fixed equations derived in the previous section are all gauge invariant. This is
a very powerful test that the equations are correct as given the complexity of the
equations it is a very non-trivial matter.

\subsection{Conformal Newtonian gauge}
The Conformal Newtonian gauge is defined by
\begin{eqnarray}
  \XiM &=& -\PsiM \\
  \chiM &=& -6\PhiM \\
   \zetaM &=& 0 \\
   \nuM  &=& 0
\end{eqnarray}
From the relation $\XiE = \XiM + \varphi$ we also set $\XiE = -\PsiE$
and from $\chiE = \chiM + 6\varphi$ we set $\chiE = -6\PhiE$.
Therefore the Einstein-Hilbert frame metric perturbations are given by
\begin{eqnarray}
  \PsiE &=& \PsiM - \varphi\\
  \PhiE &=& \PhiM - \varphi\\
  \zetaE &=& -(1-e^{-4\bar{\phi}})\alpha \\
  \nuE &=& 0
\end{eqnarray}

\subsubsection{Fluid equations}
The density constrast equation for scalar modes in the Conformal Newtonian gauge evolves as 
\begin{equation}
\dot{\delta} =  - (1+w)\left(k^2\theta -  3 \dot{\PhiM}\right)  - 3\frac{\dot{a}}{a}(C_s^2 - w)\delta
\end{equation}
where as the momentum divergence evolves as
\begin{equation}
\dot{\theta} =  -\frac{\dot{a}}{a}(1 -3w)\theta
  +\frac{C_s^2}{1+w}\delta - \frac{\dot{w}}{1+w}\theta - \frac{2}{3}\left(k^2 - \frac{3K}{r_c^2}\right) \Sigma + \PsiM 
\end{equation}

\subsubsection{Scalar field equation}
The two first order equations coming from the perturbed scalar field equation are
\begin{equation}
 \dot{\gamma} = -3\frac{\dot{b}}{b} \gamma +  \frac{\bar{\mu}}{a}e^{-3\bar{\phi}}k^2\left(\varphi 
        + \dot{\bar{\phi}}\alpha\right)
  + \frac{e^{\bar{\phi}}}{a}\bar{\mu}\dot{\bar{\phi}}\left[-6\dot{\PhiE} -2k^2\zetaE \right] 
 + 8\pi G a e^{-3\bar{\phi}}\left[ \delta\rho + 3\delta P  + (\bar{\rho} + 3\bar{P})\left(
  \PsiE - 2\varphi\right)\right]  
\end{equation}
and
\begin{equation}
   \dot{\varphi} = -\frac{1}{2U}ae^{-\bar{\phi}}\gamma +  \dot{\bar{\phi}} \PsiE 
 \label{eq:varphi_dot}
\end{equation}

\subsubsection{Vector field equation}
The scalar mode of the perturbed vector field equation is
\begin{equation}
K_B\left(\dot{E} + \frac{\dot{b}}{b}E \right) =   - \bar{\mu} \dot{\bar{\phi}}     
      (\varphi -\dot{\bar{\phi}} \alpha)  
    + 8\pi G  a^2 (1-e^{-4\bar{\phi}}) (\bar{\rho}+\bar{P}) (\theta -\alpha) 
\end{equation}
and
\begin{equation}
 \dot{\alpha}  =  E + \PsiE +  \left(\dot{\bar{\phi}} - \frac{\dot{a}}{a}\right)  \alpha 
\end{equation}

\subsubsection{Generalized Einstein equations}
The scalar modes of the perturbed generalized Einstein equations yield
for the Hamiltonian constraint
\begin{equation}
-2\left(k^2 - \frac{3K}{r_c^2}\right)\PhiE
   - 2e^{4\bar\phi}\frac{\dot{b}}{b}\left[3\dot{\PhiE} + k^2\zetaE
 + 3\frac{\dot{b}}{b}\PsiE\right] + a e^{3\varphi} \dot{\bar{\phi}} \gamma 
 - K_B k^2E 
 = 8\pi G a^2 \bar{\rho} \left[ \delta  - 2 \varphi\right]  
\label{eq:TE_den}
\end{equation}
the momentum constraint equation
\begin{equation}
\dot{\PhiE} + \frac{K}{r_c^2}\zetaE + \frac{\dot{b}}{b}\PsiE -\bar{\mu}\dot{\bar{\phi}}\varphi =
   4\pi G a^2  e^{-4\bar{\phi}}(\bar{\rho} + \bar{P}) \theta 
\label{eq:TE_vel}
\end{equation}
and the two propagation equations
\begin{eqnarray}
&&  6\ddot{\PhiE} +  2k^2\left(\dot{\zetaE} - e^{-4\bar{\phi}}\PsiE\right)
 + 2e^{-4\bar{\phi}}\left(k^2 - \frac{3K}{r_c^2}\right)\PhiE  
+ 2 \frac{\dot{b}}{b}\left[6\dot{\PhiE} + 3 \dot{\PsiE} + 2k^2\zetaE \right] 
\nonumber \\
 &&
  + 4\dot{\bar{\phi}} \left[ 3\dot{\PhiE} + k^2\zetaE \right]
   + 3\frac{\mu}{U}a e^{-\bar{\phi}} \dot{\bar{\phi}} \gamma
 -6\left[-2\frac{\ddot{b}}{b} + \frac{\dot{b}^2}{b^2} - 4\dot{\bar{\phi}}\frac{\dot{b}}{b} \right]\PsiE
  = 24\pi G a^2e^{-4\bar{\phi}}\bar{\rho} (C_s^2\delta - 2 w \varphi) 
\end{eqnarray}
and
\begin{equation}
\PhiE - \PsiE +  e^{4\bar{\phi}}\left[ \dot{\zetaE}
 + 2\left(\frac{\dot{b}}{b}+\dot{\bar{\phi}}\right)\zetaE \right]
  =    8\pi G a^2 (\bar{\rho} + \bar{P})\Sigma 
\end{equation}

\subsection{Conformal synchronous gauge}
The conformal synchronous gauge is defined by $\XiM = 0$
and $\zetaM =  0$ which fixes $\zetaE = -(1 - e^{-4\bar{\phi}})\alpha$ 
and $\XiE = \varphi$.
Following the standard notation lets also set $\chiM = h$ which gives $\chiE = h + 6\varphi$ and
$\nu = -\frac{1}{k^2}(h + 6\eta)$.

\subsubsection{Fluid equations}
The density constrast equation for scalar modes evolves as 
\begin{equation}
\dot{\delta} = - 3 \frac{\dot{a}}{a}(C_s^2 - w)\delta - (1+w)\left(
   k^2 \theta +  \frac{1}{2} \dot{h}\right)  
\label{eq:delta_dot_syn}
\end{equation}
while the momentum divergence evolves as
\begin{equation}
\dot{\theta} =  - \frac{\dot{a}}{a}(1 -3w)\theta 
  +\frac{C_s^2}{1+w}\delta - \frac{\dot{w}}{1+w}\theta 
    - \frac{2}{3} \left(k^2 -\frac{3K}{r_c^2}\right)\Sigma 
\label{eq:theta_dot_syn}
\end{equation}

\subsubsection{Scalar field equation}
The two equations equivalent to the scalar field equation are
\begin{eqnarray}
 \dot{\gamma} &=& -3\frac{\dot{b}}{b} \gamma +  \frac{\bar{\mu}}{a}e^{-3\bar{\phi}}k^2\left(\varphi 
        + \dot{\bar{\phi}}\alpha\right)
  + \frac{e^{\bar{\phi}}}{a}\bar{\mu}\dot{\bar{\phi}}\left[\dot{h} + 6\dot{\varphi} -2k^2\zetaE \right] 
\nonumber \\
 &+& 8\pi G a e^{-3\bar{\phi}}\left[ \delta\rho + 3\delta P  - 3(\bar{\rho} + 3\bar{P})\varphi\right]  
\end{eqnarray}
and
\begin{equation}
   \dot{\varphi} = -\frac{1}{2U}ae^{-\bar{\phi}}\gamma -  \dot{\bar{\phi}} \varphi
\end{equation}

\subsubsection{Vector Bekenstein equation}
The scalar mode of the perturbed vector Bekenstein equation is given by the two 
 first order equations 
\begin{equation}
K_B\left(\dot{E} + \frac{\dot{b}}{b} E \right) =   - \bar{\mu}
      \dot{\bar{\phi}}(\varphi - \dot{\bar{\phi}}\alpha) 
    + 8\pi G  a^2 (1- e^{-4\bar{\phi}}) (\bar{\rho}+\bar{P}) (\theta -\alpha) 
\end{equation}
and
\begin{equation}
 \dot{\alpha}  =  E  - \varphi +  \left(\dot{\bar{\phi}} - \frac{\dot{a}}{a}\right)  \alpha 
\end{equation}

\subsubsection{Generalized Einstein equations}
The scalar modes of the perturbed generalized Einstein equations yield
for the Hamiltonian constraint
\begin{equation}
2\left(k^2 - \frac{3K}{r_c^2}\right)\left(\varphi - \eta\right) 
   + e^{4\bar\phi}\frac{\dot{b}}{b}\left[\dot{h}  -2k^2\zetaE
 +6\frac{\dot{a}}{a}\varphi\right] + a e^{3\varphi}\left( \dot{\bar{\phi}} -\frac{3}{U} \frac{\dot{b}}{b} \right) \gamma 
 - K_B k^2E 
 = 8\pi G a^2 \bar{\rho} \left[ \delta  - 2 \varphi\right]  
\end{equation}
the momentum constraint equation
\begin{equation}
2k^2\dot{\eta} + \frac{K}{r_c^2}\left[2k^2\zetaE - \dot{h} - 6\dot{\eta} \right] 
  - 2k^2\left(\frac{\dot{a}}{a} +  \bar{\mu}\dot{\bar{\phi}} \right)\varphi
 + \frac{k^2}{U}a e^{-\bar{\phi}} \gamma  =
   8\pi G a^2  e^{-4\bar{\phi}}(\bar{\rho} + \bar{P}) k^2 \theta 
\end{equation}
and the two propagation equations
\begin{eqnarray}
&&  -\ddot{h} - 6\ddot{\varphi} +  2k^2\dot{\zetaE} 
+ \frac{6K}{r_c^2}e^{-4\bar{\phi}}\varphi
 + 2e^{-4\bar{\phi}}\left(k^2 - \frac{3K}{r_c^2}\right) \eta
- 2 \frac{\dot{b}}{b}\left[\dot{h} + 9 \dot{\varphi} - 2k^2\zetaE \right] 
\nonumber \\
 &&
  - 2\dot{\bar{\phi}} \left[ \dot{h} + 6\dot{\varphi}  - 2k^2\zetaE \right]
   + 3\frac{\mu}{U}a e^{-\bar{\phi}} \dot{\bar{\phi}} \gamma
 +6\left[-2\frac{\ddot{b}}{b} + \frac{\dot{b}^2}{b^2} - 4\dot{\bar{\phi}}\frac{\dot{b}}{b} \right]\varphi
  = 24\pi G a^2e^{-4\bar{\phi}}\bar{\rho} (C_s^2\delta - 2 w \varphi) 
\end{eqnarray}
and
\begin{equation}
 \ddot{h} + 6\ddot{\eta} - 2 e^{-4\bar{\phi}} k^2\eta -  2k^2\dot{\zetaE}
 + 2\left[\frac{\dot{b}}{b}+\dot{\bar{\phi}}\right] \left[ \dot{h} + 6\dot{\eta} - 2 k^2\zetaE \right]
  =  -16\pi G a^2 e^{-4\bar{\phi}}(\bar{\rho} + \bar{P})k^2\Sigma 
\end{equation}

\subsection{$\alpha$-gauge}
In the $\alpha$-gauge one  sets the vector field perturbation to zero, $\alpha =  0$,
which fixes $\zetaE = \zeta$.
Since setting $\alpha=0$ essentially fixes the gauge variable $\xi$, it is no longer 
allowed to set $\XiM=0$. Therefore this gauge cannot be put in synchronous form.
As a further gauge fixing condition, let $\nu = 0$,
 which can still be done, as the gauge variable $\psi$ was not fixed at that point. 

\subsubsection{Fluid equations}
The density constrast equation for scalar modes evolves as
\begin{equation}
\dot{\delta} = - 3 \frac{\dot{a}}{a}(C_s^2 - w)\delta + (1+w)\left(
   -k^2\theta -  \frac{1}{2} \dot{\chiM} + k^2\zetaM\right)  
\end{equation}
while the momentum divergence evolves as
\begin{equation}
\dot{\theta} =  - \XiM  - \frac{\dot{a}}{a}(1 -3w)\theta +\frac{C_s^2}{1+w}\delta - \frac{\dot{w}}{1+w}\theta 
    - \frac{2}{3} \left(k^2 - \frac{3K}{r_c^2}\right) \Sigma 
\end{equation}

\subsubsection{Scalar field equation}
The two equations equivalent to the scalar field equation are
\begin{eqnarray}
 \dot{\gamma} &=& -3\frac{\dot{b}}{b} \gamma +  \frac{\bar{\mu}}{a}e^{-3\bar{\phi}}k^2\varphi 
  + \frac{e^{\bar{\phi}}}{a}\bar{\mu}\dot{\bar{\phi}}\left[\dot{\chiE} -2k^2\zetaM \right] 
\nonumber \\
 &+& 8\pi G a e^{-3\bar{\phi}}\bar{\rho}\left[ (1 + 3C_s^2)\delta  - (1 + 3w)\left(E + 2\varphi\right)\right]  
\end{eqnarray}
and
\begin{equation}
   \dot{\varphi} = -\frac{1}{2U}ae^{-\bar{\phi}}\gamma -  \dot{\bar{\phi}} E 
\end{equation}

\subsubsection{Vector field equation}
The scalar mode of the perturbed vector field equation is given by the  
 first order equation 
\begin{equation}
K_B\left(\dot{E} + \frac{\dot{b}}{b}E \right) =   - \bar{\mu} \dot{\bar{\phi}}\varphi  
    + 8\pi G  a^2(1- e^{-4\bar{\phi}})(\bar{\rho}+\bar{P}) \theta 
\end{equation}

\subsubsection{Generalized Einstein equations}
The scalar modes of the perturbed generalized Einstein equations yield
for the Hamiltonian constraint
\begin{equation}
\frac{1}{3}\left(k^2 - \frac{3K}{r_c^2}\right)\chiE 
   + e^{4\bar\phi}\frac{\dot{b}}{b}\left[\dot{\chiE} -2k^2\zetaM
 +6\frac{\dot{b}}{b}E\right] + a e^{3\varphi} \dot{\bar{\phi}} \gamma 
 - K_B k^2E 
 = 8\pi G a^2 \bar{\rho} \left[ \delta  - 2 \varphi\right]  
\end{equation}
the momentum constraint equation
\begin{equation}
-\frac{1}{3}\dot{\chiE} + \frac{2K}{r_c^2}\zetaM  - 2\frac{\dot{b}}{b}E =
   8\pi G a^2  e^{-4\bar{\phi}}(\bar{\rho} + \bar{P}) \theta  + 2\bar{\mu}\dot{\bar{\phi}}\varphi 
\end{equation}
and the two propagation equations
\begin{eqnarray}
&&  -\ddot{\chiE} +  2k^2\left(\dot{\zetaM} + e^{-4\bar{\phi}}E\right)
 - \frac{1}{3}e^{-4\bar{\phi}}\left(k^2 - \frac{3K}{r_c^2}\right)\chiE 
- 2 \frac{\dot{b}}{b}\left[\dot{\chiE} + 3 \dot{E} - 2k^2\zetaM \right] 
\nonumber \\
 &&
  - 2\dot{\bar{\phi}} \left[ \dot{\chiE}  - 2k^2\zetaM \right]
   + 3\frac{\mu}{U}a e^{-\bar{\phi}} \dot{\bar{\phi}} \gamma
 +6\left[-2\frac{\ddot{b}}{b} + \frac{\dot{b}^2}{b^2} - 4\dot{\bar{\phi}}\frac{\dot{b}}{b} \right]E
  = 24\pi G a^2e^{-4\bar{\phi}}\bar{\rho} (C_s^2\delta - 2 w \varphi) 
\end{eqnarray}
and
\begin{equation}
   e^{-4\bar{\phi}}\left[2E -\frac{1}{3}\chiE \right] +  2\dot{\zetaM}
 + 4\left[\frac{\dot{b}}{b}+\dot{\bar{\phi}}\right]\zetaM 
  =    16\pi G a^2 e^{-4\bar{\phi}}(\bar{\rho} + \bar{P})\Sigma 
\end{equation}

\section{Specifics of the perturbations of the Einstein tensor}
\label{app_ein}
In this section I write out explicitely all the perturbed terms of the Einstein tensor to
help the reader to follow the calculations.

Changing connection to $\connetat_\mu$ gives
\begin{eqnarray}
 \conneta_\beta \conneta_\mu \permetE^\beta_{\;\;\nu} &=& 
 \etaE^{\alpha\beta}\connetat_\alpha\connetat_\mu\permetE_{\beta\nu} 
 + 2 e^{4\bar{\phi}}\connetat^2\bar{\phi}\left[\bar{A}^\alpha\bar{A}^\beta\permetE_{\alpha\beta}\bar{A}_\mu\bar{A}_\nu
 - \bar{A}^\lambda\bar{A}_\mu\permetE_{\lambda\nu}\right]
\nonumber \\
&&+ 2(\bar{A}^\rho\connetat_\rho\bar{\phi})\left\{ \etaE^{\alpha\beta}\bar{A}^\lambda\connetat_\alpha\permetE_{\beta\lambda}
  \bar{A}_\mu\bar{A}_\nu
 +  e^{4\bar{\phi}}\left[2\bar{A}^\alpha\bar{A}^\beta\connetat_\alpha\permetE_{\beta\nu}\bar{A}_\mu 
  - \bar{A}^\alpha\connetat_\mu\permetE_{\alpha\nu} + \bar{A}^\alpha\bar{A}^\beta\connetat_\mu
         \permetE_{\alpha\beta}\bar{A}_\nu \right]\right\}
\nonumber \\
&& + 8 e^{4\bar{\phi}}(\bar{A}^\rho\connetat_\rho\bar{\phi})^2\left[\bar{A}^\lambda\bar{A}_\mu\permetE_{\lambda\nu}
  - 2 \bar{A}^\alpha\bar{A}^\beta\permetE_{\alpha\beta}\bar{A}_\mu\bar{A}_\nu\right]
\end{eqnarray}

\begin{equation}
   \conneta_\mu \conneta_\nu \permetE^\alpha_{\;\;\alpha} = \connetat_\mu\connetat_\nu\permetE^\alpha_{\;\;\alpha} 
 + 2(\bar{A}^\rho\connetat_\rho\bar{\phi})\bar{A}_\mu\bar{A}_\nu
\bar{A}^\beta\connetat_\beta\permetE^\alpha_{\;\;\alpha}
\end{equation}

\begin{eqnarray}
 \conneta^2\permetE_{\mu\nu} &=&
 \etaE^{\alpha\beta}\connetat_\alpha\connetat_\beta\permetE_{\mu\nu} 
  - 2e^{4\bar{\phi}}\bigg\{2\connetat^2\bar{\phi} \bar{A}_{(\mu}\bar{A}^\lambda\permetE_{\nu)\lambda}
 +(\bar{A}^\rho\connetat_\rho\bar{\phi})\left[\bar{A}^\lambda\connetat_\lambda\permetE_{\mu\nu}
 -4\bar{A}^\alpha\bar{A}^\beta\connetat_\alpha\permetE_{\beta(\mu}\bar{A}_{\nu)}\right]
\nonumber \\
&& + 4(\bar{A}^\rho\connetat_\rho\bar{\phi})^2\left[ 
\bar{A}^\alpha\bar{A}^\beta\permetE_{\alpha\beta}\bar{A}_\mu\bar{A}_\nu
-2\bar{A}^\beta\bar{A}_{(\mu}\permetE_{\nu)\beta} \right]
\bigg\}
\end{eqnarray}

\begin{eqnarray}
  \conneta_\alpha \conneta_\beta \permetE^{\alpha\beta} &=&
 \etaE^{\alpha\mu}\etaE^{\beta\nu}\connetat_\alpha\connetat_\beta\permetE_{\mu\nu}
 + 2 e^{8\bar{\phi}}\bigg\{
 (\bar{A}^\rho\connetat_\rho\bar{\phi})\left[3
    \bar{A}^\lambda\connetat_\lambda\left(\bar{A}^\mu\bar{A}^\nu\permetE_{\mu\nu}\right) - 2e^{-4\bar{\phi}}\bar{A}^\lambda
       \etaE^{\mu\nu}\connetat_\mu\permetE_{\lambda\nu}\right]
\nonumber \\
 && - 2\bar{A}^\mu\bar{A}^\nu\permetE_{\mu\nu}\left[\connetat^2\bar{\phi} - 6\bar{A}^\rho\connetat_\rho\bar{\phi})^2\right]
\bigg\}
\end{eqnarray}

\begin{equation}
  \conneta^2\permetE^\alpha_{\;\;\alpha}
   = \etaE^{\rho\lambda}\connetat_\rho\connetat_\lambda\permetE^\alpha_{\;\;\alpha}
  - 2(\bar{A}^\rho\connetat_\rho\bar{\phi}) e^{4\bar{\phi}}\bar{A}^\lambda\connetat_\lambda\permetE^\alpha_{\;\;\alpha}
\end{equation}

\begin{eqnarray}
  2\tilde{f}^\alpha_{\mu\nu}\conneta_\alpha\ln b &=&
 e^{4\bar{\phi}}(\bar{A}^\lambda\connetat_\lambda\ln b)\left[
   \bar{A}^\beta\connetat_\beta\permetE_{\mu\nu}
  -  2\bar{A}^\beta\connetat_{(\mu}\permetE_{\nu)\beta}
 -  4(\bar{A}^\rho\connetat_\rho\bar{\phi})
    \bar{A}^\alpha\bar{A}^\beta\permetE_{\alpha\beta}\bar{A}_\mu\bar{A}_\nu\right]
\end{eqnarray}

\begin{equation}
 2\permetE^{\alpha\beta}\conneta_\alpha\conneta_\beta\ln b = 
  2 e^{8\bar{\phi}}\bar{A}^\alpha\bar{A}^\beta\permetE_{\alpha\beta}\left[-\connetat^2\ln b + 
2(\bar{A}^\lambda\connetat_\lambda\ln b)(\bar{A}^\rho\connetat_\rho\bar{\phi})\right] 
\end{equation}

\begin{eqnarray}
  2\etaE^{\alpha\beta}\tilde{f}^\lambda_{\alpha\beta}\conneta_\lambda\ln b &=&
   e^{4\bar{\phi}}(\bar{A}^\gamma\connetat_\gamma\ln b)\left[\bar{A}^\rho\connetat_\rho\permetE^\alpha_{\;\;\alpha}
 - 2\bar{A}^\beta\connetat_\alpha\permetE^\alpha_{\;\;\beta}\right]
\\
 &=&e^{4\bar{\phi}}(\bar{A}^\rho\connetat_\rho\ln b)\left[\bar{A}^\lambda\connetat_\lambda\permetE^\alpha_{\;\;\alpha} 
  -2\bar{A}^\rho\etaE^{\alpha\beta}\connetat_\alpha\permetE_{\beta\rho}
  + 8e^{4\bar{\phi}}(\bar{A}^\rho\connetat_\rho\bar{\phi})
       \bar{A}^\alpha\bar{A}^\beta\permetE_{\alpha\beta}\right]
\end{eqnarray}

\begin{equation}
 \conneta^2\ln b =  e^{4\bar{\phi}}\left[\connetat^2\ln b
  - 2(\bar{A}^\lambda\connetat_\lambda\ln b)(\bar{A}^\rho\connetat_\rho\bar{\phi})\right]
\end{equation}

\end{document}